\documentclass[11pt,a4paper]{article} 
\pdfoutput=1
\usepackage[no-natbib-sort]{my-jheppub}

\usepackage{amsmath, amssymb}

\usepackage{environ}
\usepackage{mathrsfs}
\usepackage{array,arydshln}

\usepackage{graphicx,epsfig}
\usepackage{epic}
\usepackage{youngtab}
\usepackage{float}
\usepackage{color}
\definecolor{maroon}{rgb}{0.8,0.3,0.}

\usepackage{slashed}

\usepackage[nodayofweek]{date time}

\usepackage{aurical}
\usepackage[T1]{fontenc}

\newcommand{\be}{\begin{equation}}
\newcommand{\ee}{\end{equation}}

\newcommand{\ads}{AdS$_5\times S^5$\ }

\newcommand{\mc}{\mathcal }
\newcommand{\Z}{\mathcal{Z}}
\newcommand{\N}{\mathcal{N}}
\newcommand{\E}{{\mathcal E}}
\newcommand{\A}{{\mathcal A}}

\def \bW  {{\cal W}}

\newcommand{\la}{\longrightarrow}

\def\XXint#1#2#3{{\setbox0=\hbox{$#1{#2#3}{\int}$}
     \vcenter{\hbox{$#2#3$}}\kern-.5\wd0}}

    \newcommand{\beq}{\begin{equation}}
    \newcommand{\eeq}{\end{equation}}
    \newcommand\bea{\begin{eqnarray}}
    \newcommand\eea{\end{eqnarray}}

\def \del{ \partial}
\def \la {\label}
\newcommand{\rf}[1]{(\ref{#1})}
\def\ov{\over}
\def\no{\nonumber} \def \aa {{\rm a}}

\def \ci {\cite}
\def \p {\phi}
 
\def \ed {\end{document}}
 \def \r {\rho} 

\def \foot {\footnote}
\def \b {\beta} 
\def \dd {{\rm d}} 
\def \om {\omega} 
\def \tr {{\rm tr}}

 \def \ha {{{1 \ov 2}}}

\def \cc {{\rm c}}
\def \aa {{\rm a} }
\def \z {\zeta}

\def \OO  {{\mc O}}   \def \tr  {{\rm tr }}   

\def \iffa  {\iffalse}

\title{
Supergravity one-loop corrections on AdS$_7$  and  AdS$_3$,  higher spins   and AdS/CFT
}

\author[a]{Matteo Beccaria,} 
\author[a]{Guido Macorini,} 
\author[b]{Arkady A. Tseytlin\footnote{Also at Lebedev Institute, Moscow}}

\affiliation[a]{Dipartimento di Matematica e Fisica Ennio De Giorgi,\\
Universit\`a del Salento \& INFN, Via Arnesano, 73100 Lecce, 
Italy} 

\affiliation[b]{The Blackett Laboratory, Imperial College, London SW7 2AZ, U.K.}

\emailAdd{matteo.beccaria@le.infn.it}
\emailAdd{macorini@nbi.ku.dk}
\emailAdd{tseytlin@imperial.ac.uk}

\abstract{
As was shown earlier,  the one-loop correction in 10d supergravity on  $AdS_5 \times S^5$ 
 corresponds to the  contributions to the   vacuum energy and 4d boundary conformal anomaly which  are  minus 
the values for one $\N=4$ Maxwell  supermultiplet, thus   reproducing the subleading term in the
$N^2-1$  coefficient in the dual $SU(N)$   SYM theory. 
We  perform  similar  one-loop computations  in  11d  supergravity on $AdS_7 \times S^4$ 
and 10d supergravity  on $AdS_3 \times S^3 \times T^4$. In the $AdS_7$   case  we find  that 
the   corrections to the 6d   conformal   anomaly a-coefficient and the vacuum energy  are again  minus 
the ones  for one (2,0) tensor multiplet, suggesting that the total a-anomaly coefficient  for the  dual (2,0) 
theory  is  $ 4 N^3 - 9/4 N - 7/4$  and thus vanishes  for $N=1$. 
In the  $AdS_3$  case   the one-loop  correction to the  vacuum energy or 
 2d central charge  turns out to be  equal to that of  one free
(4,4)   scalar  multiplet, i.e.  is $c=+6$.  This reproduces  the subleading term in the  central charge 
$c= 6(Q_1 Q_5 +1)$ of the dual 2d CFT  describing  decoupling  limit of D5-D1 system. 
We    also present the  expressions  for the 6d a-anomaly coefficient  and vacuum energy  contributions
 of    general-symmetry higher spin field  in $AdS_7$   and consider  their 
 application to tests of vectorial AdS/CFT   with the boundary conformal  6d  theory  represented  by   free 
 scalars,  spinors or  rank 2 antisymmetric tensors. 
 }
 
\iffa  
Authors:
M. Beccaria, G. Macorini, A.A. Tseytlin
Report No.:
Imperial-TP-AT-2014-08
Title:     Supergravity one-loop corrections on AdS7  and  AdS3,  higher spins   and AdS/CFT
Comments:    28 pages. 
Abstract: 
As was shown earlier,   one-loop correction in 10d supergravity on  AdS_5 x S^5
 corresponds  to the  contributions to the  vacuum energy and  boundary 4d  conformal anomaly which  are  minus 
the values for one  n=4  Maxwell  supermultiplet, thus   reproducing the subleading term in their 
N^2-1  coefficient in the dual $SU(N)$   SYM theory. 
We  perform  similar  one-loop computations  in  11d  supergravity on  AdS_7 x S^4
and 10d supergravity  on  AdS_3 x S^3 x T^4. In the  AdS_7   case  we find  that 
the    corrections to the 6d   conformal   anomaly a-coefficient and the vacuum energy  are again  minus 
the ones  for one (2,0) tensor multiplet, suggesting that the total a-anomaly coefficient  for the  dual (2,0) 
theory  is  4 N^3 - 9/4 N - 7/4  and thus vanishes  for N=1. 
In the  AdS_3  case   the one-loop  correction to the  vacuum energy or 
 2d central charge  turns out to be  equal to that of  one free
(4,4)   scalar  multiplet, i.e. is  c=+6.  This reproduces  the subleading term in the  central charge 
c= 6(Q_1 Q_5 +1) of the dual 2d CFT  describing  decoupling  limit of D5-D1 system. 
We    also present the  expressions  for the 6d  anomaly a-coefficient  and vacuum energy  for a 
   general-symmetry higher spin field  in AdS_7   and consider  their 
 application to tests of vectorial AdS/CFT   with the boundary conformal  6d  theory  represented  by   free 
 scalars,  spinors or  rank 2 antisymmetric tensors. 
\fi

\iffa 
We consider the conformal anomaly  for the boundary theory in instances of the 
AdS$_{7}$/CFT$_{6}$ and AdS$_{3}$/CFT$_{2}$
dualities. In the first case, we study the 6d (2,0) theory describing $N$ coincident M5-branes  and dual 
to M-theory on $AdS_7 \times S^4$. At large $N$, the conformal anomaly  has a subleading term $\mc O(N^{0})$ that 
is captured by the Kaluza-Klein fields arising from compactification of 11d supergravity on $S^{4}$. Using
a suitable $\zeta$-function regularisation, we show that the bulk vacuum energy and conformal anomaly
can be computed for the full tower of KK fields. The resulting values are opposite to those of a single (2,0) multiplet
in close analogy with the 4d case of duality  between $\N=4$  $SU(N)$  SYM   theory and 
string  theory in $AdS_5 \times S^5$. Also, 
we discuss applications to vectorial AdS/CFT by considering the free 6d CFT of a single (2,0) multiplet with $U(N)$
or $O(N)$ global symmetry and the duality of its singlet sector with a higher spin gauge theory in AdS$_{7}$.
In the case of AdS$_{3}$/CFT$_{2}$
duality, we consider type IIA or IIB string  theory in  $AdS_3 \times S^3 \times M^4$  space, with $M=T^{4}, K3$.
The boundary 2d CFT corresponds to the gauge theory  describing the low-energy limit  of the D5-D1 system. 
The central charge of this CFT is $c= 6(Q_1 Q_5 +1)$, where $Q_{i}$ are the number of branes. Again, 
for large $Q_{1}Q_{5}$, the subleading constant term $+6$  must be reproduced by a one-loop computation in 
10d supergravity taking into account the towers of KK modes on $S^{3}\times M^{4}$. We check this prediction 
in all cases by an explicit analysis of the relevant 
6d supergravities with supersymmetry $(\N_{L}, \N_{R})=(2,0), (1,1), (2,2)$.
\fi 

\allowdisplaybreaks

\begin{document}

\begin{flushright}\small{Imperial-TP-AT-2014-08}\end{flushright}

 \maketitle

\flushbottom
\def \De {\Delta} 
\def \ads {AdS$_{5}$\ }
\def \te {\textstyle} \def \iffa {\iffalse} 

\def \ha {{\te {1 \ov 2}}}

 \def  \ba { \begin{align} }
 \def  \ea { \end{align} }

\def \gg   {{\rm g}}
\def \cc    {{\rm c}} 
\def \aa  {{\rm a}}

\def \ep {\epsilon}
 \def \k {\kappa} \def \r {\rho} 

\def \RR {{\rm R}}
\def \OO {{\cal O}} 
\def \edd {\end{document}} 
\def \td {\tilde} 

\def \tO {{\td \OO}}\def \rZ {{\rm Z}}

 \def \Deltat  {{\bar \OO}}
 
\def \de {\delta} 

\newcommand{\hh}{{\textstyle\frac{1}{2}}}
\def \adss {$AdS_5 \times S^5$ }
\def \adsss {$AdS_7 \times S^4$ }
\def \adse {$AdS_7$ }
\def \ads {$AdS_5$ }
\def \adt {$AdS_3$ }

\section{Introduction}

One  of the key probes of  the AdS/CFT  correspondence \ci{Maldacena:1997re,Gubser:1998bc,Witten:1998qj}
is the   boundary theory conformal anomaly which is closely related to the simplest  correlators of the stress tensor 
\ci{Liu:1998bu, Henningson:1998gx, Blau:1999vz}.
In the case of the duality  between $\N=4$  $SU(N)$  SYM   theory and string  theory in $AdS_5 \times S^5$ 
the   gauge-theory result  for the Weyl anomaly  is   $\A_4 =  -\aa\, \E_4  + \cc\, \bW_4 $, \    $\aa=\cc= (N^2-1) k_1$
($k_1 = {1\ov 4}$   is the   contribution of a single $\N=4$ vector multiplet). It 
is determined by the 2- and 3-point  correlators of stress tensor and   should thus be exact.
The $N^2$  term  is indeed  reproduced  at strong coupling  by  the classical  supergravity action  \ci{Henningson:1998gx}.
 
It was suggested in \ci{Mansfield:2000zw,Mansfield:2003gs}\foot{This follows the  
 analogy with what happens in the case of  the R-symmetry anomaly \cite{Bilal:1999ph}.}
 that the   -1 term  in $N^2 -1$  coefficient  
should   come from  the one-loop 10d supergravity  correction (the contribution of all  massive string mode  multiplets 
should  vanish). 
This was   recently  confirmed in \ci{Beccaria:2014xda}
where it was found that  the contributions of  the massless 5d supergravity  modes 
and   the massive  $S^5$  KK modes  to the boundary conformal anomaly can   be universally  described 
by a simple formula: $\aa_p=\cc_p = p k_1$, where  
   $p=1$ for a  vector multiplet (or boundary doubleton to be omitted), 
$p=2$ for the massless  5d  supergravity modes, 
and $p=3,4, ...$  for the massive  KK   levels. 
Summing over $p$ using a  special   regularization prescription  $\sum_{p=1}^\infty p  =0$ 
 (which  is, in fact, required  for consistency   with    the standard 
$\zeta$-function regularization for the  Casimir energy in 10d) 
gives indeed 
$(\aa=\cc)_{\rm 1-loop \ sugra} =-1$.

Below  will    perform  a similar one-loop computation of  the boundary $\aa$-anomaly 
  in the case of  11d supergravity on $AdS_7 \times S^4$  (correcting an  earlier attempt in \ci{Mansfield:2003bg}). 
This  will  determine the   subleading   $N^0$ term  in the  $\aa$-coefficient of conformal anomaly  
of  the 6d (2,0) theory describing $N$ coincident M5-branes    which should   be dual to M-theory on $AdS_7 \times S^4$. 

In addition to the  duality examples   based on $AdS_5 \times S^5$  and $AdS_7 \times S^4$ 
supergravity backgrounds  there is also
 the duality \ci{Maldacena:1997re,Aharony:1999ti} 
   between    string  theory in  $AdS_3 \times S^3 \times T^4$  space  supported by  RR 3-form flux and 
  2d  CFT  corresponding to  gauge theory  describing low-energy limit  D5-D1 system. 
  The  central  charge of this CFT is $c= 6(Q_1 Q_5 +1)$ \ci{Witten:1997yu,Aharony:1999ti} 
  ($Q_i$ are the number of branes). The leading $6Q_1Q_5$   can be  reproduced from the classical action of  10d 
  supergravity on $  S^3 \times T^4$  \ci{Henningson:1998gx,Balasubramanian:1999re}. Here 
 we shall  demonstrate that the subleading  $+6$ term  is reproduced by  the one-loop  10d supergravity 
 contribution. This provides a non-trivial test  of this AdS$_3$/CFT$_2$ duality.

\subsection{AdS$_7$/CFT$_6$}

The conformal anomaly of a classical Weyl invariant theory in 6d has the following general form
 \cite{Bonora:1985cq,Deser:1993yx,Bastianelli:2000hi}  
\be\la{1} 
\mc A_6 = \aa\,\E_{6}+ W_6+D_6 \ , \ \ \ \ \ \  \ \ \     W_6=\cc_{1}\,I_{1}+\cc_{2}\,I_{2}+\cc_{3}\,I_{3}\ , 
\ee
where $\E_{6}$ is the Euler density in six dimensions, 
$ W_6$   is a combination of   three independent   Weyl invariants 
and $D_6$ is a total derivative  term  
(which  can be changed by adding a local counterterm and thus depends on a scheme).  
Omitting the derivative $D_6$ term,   the conformal anomaly   corresponding to a single    6d tensor multiplet 
\ci{Bastianelli:2000hi}    and the  6d  conformal anomaly contribution 
coming from the  classical 11d supergravity action on $S^7$  \cite{Henningson:1998gx}
(that should be  representing the   large $N$ limit of the (2,0)   theory result) 
may  be written as\foot{Here  $\mc A_6$  stands for the  integrand  of the Seeley coefficient $b_6$ 
up to the overall  factor $ - { 1 \ov 3^2  \times 2^5}$   required to  reconcile   the  normalized values of  $\aa$   coefficient below 
with the ones in eqs. (3.3) and  (A.2); see  also  \cite{Tseytlin:2000sf} for other notation.}
\begin{align}
\label{1.2}
\mc A_6 &=\aa \,\E_{6}+\cc\, \bW_6 \ ,\qquad \qquad \ \ \ 
\bW_6 \equiv   96  I_1 + 24  I_2 -  8 I_3  \ , 
\\
\label{1.3} \aa_{\rm tens} &=  {\te {7 \ov 4}} \ , \ \ \ \ \  \cc_{\rm tens} =   1 \ , \  \ \ \ \qquad 
\aa_{(2,0)}  = 4\,N^{3} + ...\ ,\qquad   \cc_{(2,0)} =  4\,N^{3} + ...\  .
\end{align}
The fact that the  anomaly   in these  two cases 
  contains  the same   Weyl-invariant combination $\bW_6$  (so that its Weyl-tensor  or B-anomaly part   is effectively parametrized 
  by  just  one   overall coefficient  $\cc$)
   is related to    non-renormalization of 
 the ratio of the 2- and 3- point correlation functions of the  corresponding 
 stress tensor \cite{Bastianelli:1999ab}.\foot{The 
 $\aa$-coefficient in 6d is related to 4-point stress tensor  correlator and may thus  receive  a more non-trivial renormalization.}

By analogy with  a subleading  order-$N$   term  in the  R-symmetry   anomaly of (2,0) theory 
\cite{Harvey:1998bx}  it was suggested   in \cite{Tseytlin:2000sf} that there  should be   also order $N$ contributions 
to $\aa_{(2,0)}$ and $\cc_{(2,0)}$ coming from the $R^4$   term in the  M-theory  11d effective action, 
\be 
\aa_{(2,0)}  = 4\,N^{3} -  {\textstyle \frac{9}{4}}\,N + a_{1}  \ ,\qquad \qquad    \cc_{(2,0)} =  4\,N^{3} - 3 \, N + c_1\  .
\la{1.4}
\ee
In \cite{Tseytlin:2000sf} the further  $N^0$  corrections $a_1,c_1$ were  ignored, 
while  
the coefficients of order $N$  terms  were  fixed so that  the resulting $N^3 + N$ terms interpolated to $N=1$ 
match the single  tensor-multiplet   anomalies in \rf{1.3}.  
As in the  case of 10d supergravity on $S^5$, one may expect 
 that $a_1$ and $c_1$     should be  determined   by the one-loop  11d supergravity correction 
 \ci{Mansfield:2003bg}.
 
Following the example of  the  D3-brane-based  \adss duality  where  the full anomaly coefficient  $N^2 -1$ 
vanishes for $N=1$   it is natural to expect that here  too the boundary  singleton (single M5-brane  tensor multiplet) 
should decouple and thus  the full 6d anomaly of the (2,0) theory 
should  vanish for $N=1$.  
This   suggests that  $a_1$ and $c_1$   should  be  non-zero 
and given by  minus the values for a single  tensor multiplet in \rf{1.3}
\be \la{15} 
a_1 = - \aa_{\rm tens} =  -{\te {7 \ov 4}} \ , \qquad \qquad 
c_1 = - \cc_{\rm tens} =  -{1 } \ .
\ee 
It was  noted in \cite{Beem:2014kka} that  the expression 
   $\cc_{(2,0)} =  4\,N^{3} - 3 \, N -1= (N-1) ( 2 N+1)^2 $  is exactly the 
    same as  the central   charge of the $A_{N-1}$  Toda  theory at the ``symmetric'' coupling point 
    (cf. also 
    \ci{Alday:2009aq,Bonelli:2009zp}).\foot{6d CFT  with $(2,0)$ supersymmetry 
   possess a protected sector of 
operators and observables  related  to a 2d  chiral algebra
 \cite{Beem:2014kka}  which is 
$\mc W$-algebra  labelled by a  simply-laced Lie algebra $\mathfrak g$ 
for a specific value of the central charge. In the $\mathfrak g=A_{N-1}$ case 
this   leads  to  $c_1=-1$.} 

Here we shall  provide support   for \rf{15} 
by   showing that  the  one-loop 11d supergravity correction indeed produces   the  value 
 $a_1=- \aa_{\rm tens}$.    Then  the expected exact value of $\aa_{(2,0)}$ is\foot{The non-vanishing 1-loop supergravity correction to the  conformal anomaly implies  that there should be also a 
similar correction  also to the  corresponding 
 R-symmetry  anomaly    (i.e. $N \to N-1$ in the $I_8$ term in the anomaly \ci{Witten:1996hc}) 
 implying its vanishing for $N=1$.  The chiral anomaly of the boundary theory is accounted for 
by  the Chern-Simons terms in the supergravity action. 
In the case of  $AdS_{5}\times S^5 $ the 
1-loop supergravity correction  shifts the Chern-Simons coefficient  $N^{2}\to N^{2}-1$ \cite{Bilal:1999ph}.
A similar  shift is  then expected in the  $AdS_{7} \times S^7$ case 
where  the CS term reproduces  the leading $N^{3}$ anomaly and also the $\mc O(N)$ correction  \cite{Harvey:1998bx}.} 
 \be 
 \te \aa_{(2,0)}  = 4\,N^{3} -   \frac{9}{4}  \,N  -   {7 \ov 4}  = ( N-1) \big( 4 N^2 + 4 N  + { 7 \ov 4} \big) \ .\la{16}\ee

Below we    shall  consider the one-loop  
 11d supergravity on $S^7$  supergravity  contributions  in the case when the 6d boundary of $AdS_7$  is 
 either  $S^6$  
  (determining   the  a-anomaly part of $\A_6$)  or  $R  \times S^5$ 
 (finding  the  vacuum or   Casimir energy $E_{c}$).
We will find that  in both  cases  the result is minus that of a single tensor multiplet
\be
\label{1.8}
\aa_{1-\rm loop\ sugra } = -\aa_{\rm tens.}, \qquad \qquad E_{c\, }{}_{1-\rm loop \ sugra} = -E_{c\, }{}_{\rm tens.} \ . 
\ee
 We  shall  use  similar  methods as  in the  $AdS_5 \times S^5$   case in  \cite{Beccaria:2014xda},
 i.e.  first  determine the contributions to $\aa$ and $E_c$  coming  from a generic $AdS_7$   higher spin  field in 
 representation $(\De; h_1,h_2,h_3)$   of $SO(2,6)$  and then  sum up  the contributions  of the relevant fields
 appearing in the supergravity    spectrum.
 
 We shall also apply our general   expressions  for $\aa(\De; h_1,h_2,h_3)$  and  $E_c(\De; h_1,h_2,h_3)$
 to provide tests  of the vectorial  AdS/CFT duality  \ci{Giombi:2013fka,Giombi:2014iua,Giombi:2014yra} 
  in the case when the  boundary 
 theory  is represented by a  free scalar, spinor or  tensor  singleton.

\subsection{AdS$_3$/CFT$_2$}

The  2d CFT  dual to  superstring in $AdS_3 \times S^3 \times T^4$  with  RR charges $Q_5,Q_1$ 
is described by a  coupled system of  three  (4,4) supersymmetric 
multiplets (see \ci{Witten:1997yu,Aharony:1999ti} and
\ci{Sax:2014mea} for a recent review): $U(Q_1)$  adjoint vector multiplet, $U(Q_1)$  adjoint  hypermultiplet, 
and  $U(Q_1)\times U(Q_5) $ bi-fundamental  hypermultiplet. The contribution to 2d conformal 
anomaly of  a single free  (4,4)  hypermultiplet (with 4 real scalars and 4 real fermions) is 
$c=4 + 4 \times \ha  =6$.\foot{In 2d  the conformal anomaly is ${\cal A}_2 = 4 \pi b_2 = \aa R, \ \ 
\aa= { 1 \ov 6 } c,$ so that $c=1$  for one real scalar.}
 The  2d vector multiplet has an  irrelevant kinetic term and thus contributes  to anomaly  only through 
measure (or ghost) factor, with single $U(1)$ vector   giving negative contribution  $c=-1$.\foot{The 
contribution of  ``non-dynamical''  2d vector gauge  field to the central charge is negative (-1)  \ci{Fradkin:1981dd}
 just like that of  non-dynamical  2d gravity (-26) \ci{Polyakov:1981rd}.
The reason  for  this -1 contribution can be  understood  also 
by   giving  vector a mass  by coupling it to a complex scalar so that it will not contribute to $c$; 
then  the  central charge of the scalar part  is reduced by 1 as  one scalar component is absorbed  by  the vector.
}
The $U(1)$  part of the  vector multiplet   is decoupled (representing the c.o.m. of the bound D5-D1 system)  
   and thus 
the total   central charge count is\foot{The same result is found  by counting  the $SU(2)$ chiral anomaly  
of  the (4,4)  superconformal algebra \ci{Witten:1997yu,Aharony:1999ti}.}
\be \la{19}
c= 6 Q_1 Q_5  +  Q_1^2    - 6 (Q_1^2-1) = 6 Q_1 Q_5 + 6     \ , \ee
where the first term is the contribution of bi-fundamental hypers,  the second -- of adjoint hypers 
and the third one of the vectors (with the $U(1)$  part subtracted).\foot{The (4,4) vector multiplet  contains 
 one 2d vector $A_m$, 4  scalars $\p_i$, 4 real spinors $\psi_k$  and 3 auxiliary fields $D_r$, all having 
 canonical dimensions (i.e. 1 for $A_m$ and $\phi_i$, $\ha$  for $\psi_k$ and 2 for $D_r$). With these dimension assignments
  the corresponding 2d conformal anomaly  can be  found from the following dimensionless  action
 (same as the standard one but with  each  kinetic term contianing  an extra $\del^{-2}$   factor)
 $\int d^2 x  \big[ ( \A^\perp_m)^2 + \phi_i^2 +  D_r \del^{-2} D_r +  \psi_k \del^{-1} \psi_k  \big]$. As as a result, the total 
 central charge   contribution is  $c= -1 + 0 + 3 \times (-1)   +  4 \times (-\ha ) =-6$.}
 
A peculiarity  of the 2d case is that  here  the subleading 
(for large $Q_5$)  term  in the central charge  which  is responsible for subtraction of the decoupled c.o.m.  modes 
enters with plus rather than  minus   sign (as was in 4d and 6d  examples).  Still, we shall  demonstrate 
 below  that as 
in the $AdS_5$ and $AdS_7$ cases   this extra $+6$ term (which should be protected and thus  receive contributions only from the BPS modes)  is also   reproduced  on the dual  AdS theory side by the  corresponding 
one-loop   correction in   10d supergravity on  $AdS_3 \times S^3 \times M^4$ with $M^4=T^4$ or $K3$. 

More precisely, instead of computing directly the correction  to the central $c$  we shall determine 
 the  one-loop 
correction to the $AdS_3$ vacuum energy or  $S^1$ Casimir energy in 2d; the latter    should be   directly related to the central charge 
\ci{Bloete:1986qm}
\be  \la{20}
\te E_c = - {1 \ov 12} \, c   \ , \ \ \ \ \  \ \  {\rm i.e.} \ \ \ \ \ \ \ \ \   c=6\ \  \ \leftrightarrow  \ \ \ \   E_c= - {1\ov 2}   \ . \ee
We shall   find   that  the one-loop supergravity contribution gives indeed 
$E_c= - {1\ov 2}$  after summing over the contributions  of  the 
KK modes  of  10d supergravity   on $ S^3 \times M^4$.

  \


The rest of this  paper is organized as follows.
In   section 2 we shall present  the expressions for the a-anomaly coefficient and  the vacuum energy 
of a    higher-spin field in \adse  corresponding  to an  arbitrary (massive or massless) representation of $SO(2,6)$, 
generalising earlier results for symmetric tensors to  mixed  symmetry case.  

In section 3 we shall  apply these results to compute  the one-loop corrections  to the 6d boundary a-anomaly and vacuum energy 
in 11d supergravity  compactified on $S^7$   obtaining eq. \rf{1.8}. 
As another application, in section 4  we shall perform  checks of vectorial AdS$_7$/CFT$_4$   duality   in the cases when  the boundary 6d theory is represented  by free scalars, spinors or (self-dual)  rank 2 tensors. We shall  find that matching of both a-anomaly and Casimir energy 
requires  particular shifts of the inverse coupling of the \adse higher spin theory. 

In section 5 we shall  turn to the case of  10d  supergravity  in $AdS_3 \times S^3 \times M^4$  and compute  
the corresponding one-loop   correction to the vacuum energy, demonstrating  that it is equal to $- {1 \ov 2}$  as in \rf{19},  thus deriving the subleading term in the central charge  \rf{20} on the dual string theory side. 

There are several technical appendices. 
In appendix A we      present the   expressions   for 
 the Casimir energy, a-anomaly and partition function  for 
 the fields of  the free $(2,0)$ multiplet in 6d. 
In Appendix B we  derive  the 6d boundary 
a-anomaly coefficient  corresponding  to  a  generic  higher spin field on $AdS_{7}$ using spectral $\zeta$-function
method.
Appendix C collects    decompositions of tensor products of  two $SO(2,6)$ singleton 
representations with spin $0, \hh, 1$ into infinite sums of  other representations  and the corresponding relations for the characters.
These Flato-Fronsdal like relations are  used  in the discussion of
applications to vectorial AdS/CFT duality in section 4. 
Appendix D   contains discussion of some properties of 
the Casimir energy of spin $0, \hh, 1$ singletons in  $AdS_{d+1}$    for general $d$. 
They are useful in comparing the 6d results to the   previously studied 4d case. 
In appendix E we list  the explicit field content of the 
$SU(2,2|1)\times SU(2,2|1)$ building blocks appearing in the Kaluza-Klein towers of 6d supergravity compactified on $S^3$. 
Appendix F  contains  the discussion of the relation  between  the expression  for the 
2d Casimir   energy  in section 5  and the  2d central  charge  derived \ci{Giombi:2013yva}  using $AdS_3$ method
for  short $SU(2,2|1)\times SU(2,2|1)$   multiplets. 



\section{Casimir  energy and $\aa$-anomaly   for generic higher spin    fields  in      $AdS_{7}$}

Given a generic conformal field in 6d   we may associate to it a   field  in \adse  correspoonding to the same representation of  $SO(2,6)$.  That allows to interpret   the one-loop  contributions  for a  field in \adse  in terms of 
Casimir  energy and   conformal anomaly of the boundary field  (see  \cite{Beccaria:2014xda}  and refs. there). 

The $SO(2,6)$  conformal group  representations will be denoted 
as  $(\Delta;\, \mathbf{h})$ where $\mathbf{h} = (h_{1},h_{2},h_{3})$  are the 
 $SO(6)$ highest weights or Young tableu  labels ($h_i$ are all integers or  all half-integers
with $h_{1}\ge h_{2}\ge |h_{3}|$).\footnote{An alternative is to use $SO(6)$ Dynkin labels $[r_1,r_2,r_3]
=  (h_{2}-h_{3}, h_{1}-h_{2}, h_{2}+h_{3})$.}

The unitary irreducible representations of $SO(2,6)$ have (see, e.g., \cite{Metsaev:1995re,Dolan:2005wy})
\be\la{21} 
\begin{split}
(i) \ \ \Delta & \ge \Delta = h_{1}+4, \ 
        \quad {\rm for}\  
        h_{1}>h_{2} \geq |h_3|,            \\
(ii) \ \ \Delta & \ge \Delta = h_{1}+3, \  
\quad {\rm for}\ h_{1}  = h_2  > |h_{3}|, \\
(iii) \ \  \Delta & \ge \Delta = h_{1}+2, \ 
\quad {\rm for}\ h_{1} = h_{2} = \pm h_{3}, \\
(iv) \ \ \Delta & \ge 2 \ {\rm or}\  \Delta=0\ \quad \ \   {\rm for}\  h_1=h_2=h_3= 0.
\end{split}
\ee
 If $\Delta$ does not saturate the above inequalities
then  the character of the 
 corresponding  {\it massive}   representation
   is\foot{The   label $^ + $ indicates that this will 
represent the partition function of the corresponding \adse  field with standard (Dirichlet) boundary conditions. 
Same  quantity without $^+$   corresponds to associated conformal  field in boundary theory (see \cite{Beccaria:2014xda}  for details). 
$ \widehat { }\ $  indicates massive  representation character.}
\be
\label{3.2}
\widehat{\Z}^{+}(\Delta;\,\mathbf{h}) = \dd(\mathbf{h})\,\frac{q^{\Delta}}{(1-q)^{6}}\ ,
\ee
where $\dd(\mathbf{h})$ is the multiplicity of the   representation 
\be\la{23}
\dd(\mathbf{h}) =\frac{1}{12}(1+h_{1}-h_{2})(1+h_{2}-h_{3})(1+h_{2}+h_{3})(2+h_{1}-h_{3})
(2+h_{1}+h_{3})(3+h_{1}+h_{2}).
\ee
If  $\Delta$ is at one of the   unitarity bounds 
 the  corresponding representation is short  or {\it massless}  (i.e. corresponds  to a massless
field in \adse space)\foot{In general 
 \ci{Metsaev:1995re},  given a field in $AdS_{d+1}$   (with even $d$) corresponding to  $SO(2,d)$  representation 
$(\Delta; h_1, h_2, ...,h_{d\ov 2})$  where first  $k=0,1,2, ...$  raws of the  $SO(d)$ Young tableu may be equal, i.e. 
  $h_1=...=h_k > h_{k+1}\ge h_{k+2}\ge ...\ge h_{d\ov 2} $,  
this  field is massless if $\De= h_k - k  +d-2$.    In the  case of \rf{21}  where $d=6$ the  lower bounds in (i),(ii) and (iii) 
 correspond to $k=0,1,2$.} 
  and  its  
character  requires  a    proper subtraction of null states and their descendants.
For  the    $\Delta = h_{1}+4$  case in (i) in \rf{21} we have the following massless representation
character 
\be\la{24} 
\begin{split}
 {\Z}^{+}(h_{1}+4;\, h_1, h_2,h_3) = \widehat{\Z}^{+}(h_{1}+4;\, h_1,h_2,h_3)-
\widehat{\Z}^{+}(h_{1}+5;\,h_{1}-1,h_{2},h_{3}) \ , 
\end{split}
\ee
where $ {\Z}^{+}$ is  given in \rf{3.2}. 
For the massless  $\Delta = h_{1}+3$  case  with $h_1=h_2=h> |h_{3}|
$  in (ii) we get 
\be\la{255} 
\begin{split}
 {\Z}^{+}(h+3;\,h, h, h_{3}) &= \widehat{\Z}^{+}(h_1+3;\,h,h,h_{3})
 -\widehat{\Z}^{+}(h+4;\,h,h-1,h_{3})   \\
 &\qquad \qquad\qquad   +\widehat{\Z}^{+}(h+5;\,h-1,h-1,h_{3})\ . \end{split}
\ee
In the  massless  case of (iii)   with  $\De=h+2$    and  $\mathbf{h} = (h,h,\pm h)$  which 
  corresponds to the 
 {\em singleton} representation   the  character is 
\be\la{25} 
\begin{split}
&\Z^{+}(h+2;\,h,h,\pm h) = \widehat{\Z}^{+}(h+2;\,h,h,\pm h)
-\widehat{\Z}^{+}\big(h+3;\,h,h,\pm(h-1)\big) \\
&\quad +\widehat{\Z}^{+}\big(h+4;\,h,h-1,\pm(h-1)\big)
-\widehat{\Z}^{+}\big(h+5;\,h-1,h-1,\pm(h-1)\big).
\end{split}
\ee
In particular, it is possible to view the $(2,0)$  tensor multiplet as  supersingleton  \cite{Gunaydin:1984wc}
which is a combination of  6d   singletons with $h=0, \ha,1$: the  one-particle partition functions 
for a scalar $\phi$, Majorana-Weyl fermion $\psi$  and self-dual tensor $T$ are the 
characters of the corresponding singleton representations  (see  also  Appendices A and C) 
\ba\la{26}
&\Z_{\phi}=\Z_{\{0\}} = \Z^{+}(2;\,0,0,0) \ , \qquad
\Z_{\psi}=\Z_{\{{1\ov 2} \}}= \Z^{+}({\textstyle\frac{5}{2}};\,\hh,\hh,\hh) \ , \no \\
& \qquad\qquad \qquad \qquad 
\Z_{T} =\Z_{\{1\}}= \Z^{+}(3;\,1,1,1)  \ .
\end{align}

\def \bh {{\bar h}} \def \bfh {{\bf{h}}}

From one-particle partition function  $\Z(q)$  given by the  corresponding $SO(2,6)$  character one can extract the expression for the Casimir energy  $E_c$  as  \ci{Gibbons:2006ij}
\ba
 \label{2.7}
 &E_c= \ha \, (-1)^{F} \sum_n  \dd_n\, {\om_n}  = \ha (-1)^{F}\,\zeta_E (-1) \ , \ \ \ \ \ \ \  \\
 \label{2.8}
 & \zeta_E (z) =\sum_n  {\dd_n \ov {\om^z_n}}   =  {1\ov \Gamma(z) } \int^\infty _0 d \beta \, \beta^{z-1} \, \Z(e^{-\b}) \ .  
\end{align}
For a  generic massive  representation $(\Delta;\, \mathbf{h}) $ 
 with the character \rf{3.2}  the  corresponding Casimir energy
is found to be ($\bh \equiv h_1 + h_2 + h_3$)
\be
\label{3.8}
\widehat E_{c}^{+}(\Delta;\, \mathbf{h}) =  \frac{(-1)^{2\bh}\dd(\mathbf{h})\,}{120960}\,
(\Delta-3)\, \Big[
12\,(\Delta-3)^{6}-126\,(\Delta-3)^{4}+336\,(\Delta-3)^{2}-191
\Big].
\ee
 The expression for the a-anomaly can be found  from the one-loop partition function  on euclidean \adse 
 as explained in appendix ~\ref{app:zeta} 
\ba\no 
& \widehat\aa^+(\Delta;{\bf{h}}) = \frac{(-1)^{2\bh} \dd(\bfh)}{2\times96\times37800} \, (\Delta -3)
\Big[15 (\Delta -3)^6 \\
&  \qquad \qquad \qquad   -21 (\Delta -3)^4 \left[h_3^2+h_1 \left(h_1+4\right)+h_2 \left(h_2+2\right)+5\right]\no 
\\ \no 
&   \qquad \qquad \qquad 
 +  35 (\Delta -3)^2 \big[\left(h_1+2\right)^2
   \left(h_2+1\right)^2+\big(h_1 \left(h_1+4\right)+h_2 \left(h_2+2\right)+5\big) h_3^2\big] \\
   &    \qquad \qquad \qquad   
   - 105  \left(h_1+2\right)^2
   \left(h_2+1\right)^2 h_3^2\Big] \ .  \la{210}
\end{align}
In the case of 
 short representations saturating a unitarity bound  one needs to  
 combine the massive representation expression  as in   \rf{24},\rf{255},\rf{25}. 

In the  special  case of   the totally symmetric massive spin $s$ tensor representation   with 
  $\mathbf{h}=(s,0,0)$, the expression 
(\ref{210}) can be written in the  following  alternative form
\be\la{211}
\begin{split}
\widehat\aa^{+}(\Delta;\, s,0,0) &= \frac{5\, (s+2)(s+3)!}{8\,(6!)^{2}\,\pi\, s!} \\
&\qquad  \times \int_{3}^{\Delta}dx\,
(x-3)(x+s-1)(x-s-5)\Gamma(x-1)\Gamma(5-x)\sin(\pi x),
\end{split}
\ee
which  is in  agreement with  the earlier result in  \cite{Giombi:2013yva,Giombi:2014iua}.

\section{One-loop  correction to  
vacuum   energy and $\aa$-anomaly in 11d    supergravity  on $AdS_7 \times S^4$}
\label{sec:ads7}

Let us  now  apply the above results \rf{3.8} and \rf{210}  to compute the 
corresponding total contribution of  the fields  in the  spectrum of 11d    supergravity 
compactified on $S^4$.  The corresponding KK    spectrum  \ci{Casher:1984ym,Gunaydin:1984wc,vanNieuwenhuizen:1984iz} is given  
in Table \ref{T1} (see also \ci{D'Hoker:2000vb}).
The massless level $p=2$  correspond to the fields of  maximal gauged  7d supergravity 
with \adse  vacuum.
\begin{table}[H]
\be
\begin{array}{|c|l|c|}
\hline
& (\Delta; h_{1},h_{2},h_{3}) & USp(4) \\
\hline
&(2 p;0,0,0) & [0,p] \\
&(2 p+\frac{1}{2};\frac{1}{2},\frac{1}{2},-\frac{1}{2}) & [1,p-1] \\
&(2 p+1; 1,1,-1) & [0,p-1] \\
p\ge 2&(2 p+1; 1,0,0) & [2,p-2] \\
&(2 p+\frac{3}{2};\frac{3}{2},\frac{1}{2},-\frac{1}{2}) & [1,p-2] \\
&(2 p+2;2,0,0) & [0,p-2] \\
\hline
\hline
&(2 p+\frac{3}{2};\frac{1}{2},\frac{1}{2},\frac{1}{2}) & [3,p-3] \\
p\ge 3 &(2 p+2;1,1,0) & [2,p-3] \\
&(2 p+\frac{5}{2}\frac{3}{2},\frac{1}{2},\frac{1}{2}) & [1,p-3] \\
&(2 p+3;1,1,1) & [0,p-3] \\
\hline
\end{array}
\ \ \ 
\begin{array}{|c|l|c|}
\hline
& (\Delta; h_{1},h_{2},h_{3}) & USp(4) \\
\hline
&& \\
&& \\
&(2 p+2;0,0,0) & [4,p-4] \\
&(2 p+\frac{5}{2};\frac{1}{2},\frac{1}{2},-\frac{1}{2}) & [3,p-4] \\
p\ge 4 &(2 p+3;1,0,0) & [2,p-4] \\
&(2 p+\frac{7}{2};\frac{1}{2},\frac{1}{2},\frac{1}{2}) & [1,p-4] \\
&(2 p+4;0,0,0) & [0,p-4] \\
&& \\
&& \\
&& \\
\hline 
\end{array}
\nonumber
\ee
\caption{$SO(2,6)\times USp(4)$ representations of the fields  of 11d supergravity on 
$AdS_{7}\times S^{4}$.} 
\label{T1}
\end{table}
Contributions  of the \adse  fields  should be summed with  multiplicities corresponding to their 
$USp(4)=SO(5)$  representations.\footnote{The dimension of the  $USp(4)$   representation $[a,b]$  
 ($a,b$   are Dynkin labels) is 
$\dim(a,b) = \frac{1}{6}\,(a+1)(b+1)(a+b+2)(a+2b+3)$.}

Using \rf{3.8}  to   sum of the  vacuum  energy contributions at each level $p$  we find 
\be\la{31}
\begin{split}\te 
E_{c, p=2}^{+} = -\frac{325}{384}, \qquad E_{c, p=3}^{+} = -\frac{925}{384}, \ \ \ \ \ \ \ \ \ 
E_{c, p\ge 4}^{+} = -\frac{25}{384}\,(6\,p^{2}-6\,p+1) \ . 
\end{split}
\ee
The value for  the massless multiplet $p=2$ is in agreement with \cite{Gibbons:2006ij}.
The expressions for  the a-anomaly are similar
\be\la{32}
\begin{split}\te 
\aa_{p=2}^{+} = -\frac{91}{1152}, \qquad \aa_{ p=3}^{+} = -\frac{259}{1152}, \qquad 
\aa_{ p\ge 4}^{+} = -\frac{7}{1152}\,(6\,p^{2}-6\,p+1).
\end{split}
\ee
Recalling that  for one (2,0)  tensor multiplet (see appendix A) 
\be \la{33} \te   E_{c,\,\rm tens.} = E_{c, 1}^{+} = -\frac{25}{384} 
\ , \qquad \qquad  \aa_{\, \rm tens.} =\aa_1^+= -\frac{7}{1152} \ . \ee
we   observe that, remarkably,  both  the vacuum energy and a-anomaly  
has the following universal expressions for any  value of $p=1,2,3,...$  
\be
\label{34}
E_{c, p}^{+} = (6\,p^{2}-6\,p+1)\,E_{c,\, \rm tens.}\ , \qquad\qquad 
\aa_{p}^{+} = (6\,p^{2}-6\,p+1)\,\aa_{\, \rm tens.}\ .
\ee
This is the direct analog  to what was   found in the case of 10d supergravity 
on $AdS_5\times S^5$ in \cite{Beccaria:2014xda}   where  the role of tensor multiplet was 
played by  $\N=4$  vector one (or  superdoubleton)  and  instead of the  coefficient 
$6\,p^{2}-6\,p+1$ we had simply $p$.\foot{For comparison, in the case of 11d   supergravity on 
$AdS_4 \times S^7$  one finds \ci{Gibbons:1984dg, Inami:1984vp} that  the  contributions 
to the $AdS_4$  vacuum energy  sum up to zero at each level $p$  separately, i.e. $E_{c, p}^{+}  =0$.
The boundary conformal anomaly  also  vanishes as  the  boundary is 3-dimensional.} 

To sum over $p$  we shall  use  the same    prescription as in 
\cite{Beccaria:2014xda}, i.e. introducing a sharp cutoff   and dropping all divergent terms\foot{Explicitly, $
 \sum_{p=1}^{P} (6p^{2}-6p+1) =2P^3-P \to 0$.}
\be 
\sum_{p=1}^{\infty} (6p^{2}-6p+1) =0 \ .
 \la{35}\ee
This prescription can be justified  by 
using the spectral $\z$-function regularization directly in 11d, i.e.  before  explicitly  expanding 
in modes of $S^4$ (see below);  it is    such a  regularization that should be consistent with  diffeomorphism symmetry of 11d theory. 
\iffa Following the approach of \cite{Beccaria:2014xda}, the sum over $p$ can be done by introducing a cut-off $P$ and dropping all singular terms. This gives \footnote{An equivalent procedure is to weight each term in the sum by the factor $z^{-p}$ and taking the finite part as $z\to 1$. In our specific case, this again gives zero, as in (\ref{4.4}), because
\be
\notag
\sum_{p=1}^{\infty}z^{-p}\,(6p^{2}-6p+1) = \frac{z^2+10 z+1}{(z-1)^3} = 
\frac{12}{(z-1)^3}+\frac{12}{(z-1)^2}+\frac{1}{z-1},
\ee has zero finite part.  }
\fi

Assuming \rf{35}  we conclude that  if the boundary  (2,0) singleton were  included in the spectrum 
of 11d supergravity,  the total vacuum energy and a-anomaly   would vanish. However, it should 
be left out representing  gauge  degrees of freedom. Thus we conclude that 
the  total one-loop supergravity  contributions are exactly 
 minus  the tensor multiplet ones
\be
\label{36}
\sum_{p=2}^{\infty}E_{c, p}^{+} = -E^{+}_{c,1} = -E_{c,\, \rm tens.}\ , \qquad\qquad 
\sum_{p=2}^{\infty}\aa_{p}^{+} = -\aa^{+}_{1} = -\aa_{\, \rm tens.} \ , 
\ee
as claimed in  (\ref{1.8}). 

Let us now  demonstrate that the prescription \rf{35}   is indeed equivalent  to the use of spectral $\zeta$-function  directly  in 11d theory. We shall   consider the   case of the Casimir energy 
  (for a similar discussion on 10d case  see    \cite{Beccaria:2014xda}). 
 For a massive 7d field in  representation $(\Delta; \mathbf{h})$ the 
vacuum energy can be  extracted from the partition function (\ref{3.2}) that we may  write in the form 
\be\la{37}
\Z^{+}(\Delta; \mathbf{h}) = \dd(\mathbf{h})\,\sum_{n=0}^{\infty}\textstyle\binom{n+5}{5}\,q^{\Delta+n}.
\ee
Then from  (\ref{2.7}),\rf{2.8}  we obtain a formal (divergent) expression for $E_{c}$ 
\ba
\label{4.7}
\widehat E^+_{c}(\Delta;\; \mathbf{h}) = \sum_{n=0}^{\infty}  e_n(\Delta;\; \mathbf{h}), \qquad \qquad 
e_n (\Delta;\; \mathbf{h}) =
\frac{1}{2}\,(-1)^{2\,\bh}\, \dd(\mathbf{h}) \,
\textstyle \binom{n+5}{5}\,(\Delta+n)\  .
\end{align}
This  sum can be   computed   using  the  $\zeta$-function  regularization applied   to
the full effective energy  eigenvalue $\Delta+n$, or, equivalently, by  introducing 
an  exponential cutoff  via 
$
e_n \   \to \ \ e_n \, e^{- \epsilon (\Delta +n)} \ ,  
$
doing  the sum, expanding  in $\ep \to 0$, and finally dropping  all singular terms.
Keeping $\ep$ finite we  may  find  the   contribution  to the 
sum \rf{4.7} from all KK states (taking into account that $p=2$ states are massless, cf. \rf{24},\rf{255}).
Denoting the   total  summand  from level $p$ as $e_n(p;\ep)$ and, summing   over  both $n$ and $p=1,2,...$, 
we obtain 
\ba
&\sum_{p=1}^{\infty}\sum_{n=0}^{\infty} e_{n}(p; \ep) =
\frac{e^{2 \epsilon }}{\left(e^{\epsilon
   /2}-1\right)^3 \left(e^{\epsilon /2}+1\right)^{11} \left(e^{\epsilon
   }+1\right)^5}
 \Big(20 e^{\epsilon /2}+50 e^{\epsilon }+100 e^{3
   \epsilon /2}+178 e^{2 \epsilon }\no \\
   &\qquad \qquad +260 e^{5 \epsilon /2}+343 e^{3
   \epsilon }+400 e^{7 \epsilon /2}+428 e^{4 \epsilon }+400 e^{9
   \epsilon /2}+343 e^{5 \epsilon }+260 e^{11 \epsilon /2}\no \\
   &\qquad \qquad +178 e^{6
   \epsilon }+100 e^{13 \epsilon /2}+50 e^{7 \epsilon }+20 e^{15
   \epsilon /2}+5 e^{8 \epsilon }+5\Big) =\te  \frac{785}{2048 \epsilon ^3}+\mc O(\epsilon).
  \end{align}
Thus  the finite part of the sum over $p\geq1$  vanishes in agreement with \rf{35}.
Equivalently, 
\be
\sum_{p=2}^{\infty}\sum_{n=0}^{\infty}e_{n}(p; \epsilon) = -\sum_{n=0}^{\infty}e_{n}(1; \epsilon) 
 \te   = -\frac{5}{16\,\epsilon^{2}}+\frac{25}{384}+\dots \ , 
\ee
in agreement with  \rf{33},(\ref{36}).


\section{Vectorial  AdS$_7$/CFT$_6$ duality}

As in lower dimensions, we may   start  with a free  CFT in 6d  described, e.g., by  $N$ (complex or real)
scalars,  spinors or 
rank 2 antisymmetric tensors
and consider the  duality  between its singlet sector   represented  by  the corresponding bilinear conserved currents  and  
higher spin  theory in  \adse  (see, e.g.,     \cite{Giombi:2014iua}). 
The   representation content  of the 7d theory is determined  from  the  Flato-Fronsdal
  type  decomposition of the product of 2 singleton representations  into sum of   higher-spin $SO(2,6)$ representations  described in appendix  
\ref{app:singleton}  (see also   \cite{Beccaria:2014xda}). 
Then using the  general  expressions for   the Casimir energy \rf{3.8}  and a-anomaly coefficient \rf{210}  
given   in section 2  we may  study  the matching of these  quantities  on the two sides   of the duality. 
In what  follows  we shall  denote by $K^{+}$ the two quantities $\aa^{+}$ and $E_{c}^{+}$  corresponding to \adse field in 
a generic massless $SO(2,6)$ representation 
 and  also use  $K=-2\,K^{+}$ for the associated  boundary conformal field values.

Starting  with the case of a    free conformal scalar   boundary 6d theory,  
the  corresponding   fields of the dual \adse theory  (``type A'' theory) 
are 
 massless totally symmetric  tensors  with spin $s$,  for which we find  from   \rf{3.8},\rf{210}    
\ba
 &\te  E_{c}^{+}(s+4;\, s,0,0) = -\frac{1}{483840}\,\nu^{2}\,(12\,\nu^{3}-58\,\nu^{2}-6\,\nu+117)\
\la{41}\ , \ \   \nu\equiv (s+1)(s+2)  \\
&\te  \aa^{+}(s+4;\, s,0,0) = -\frac{1}{29030400}\,\nu^{2}\,(22\,\nu^{3}-55\,\nu^{2}-4\,\nu+2) \ . 
 \la{42}
\end{align}
The Casimir energy \rf{41}  is a simple
extension of the results in  \cite{Giombi:2014yra}. 
The a-anomaly expression \rf{42}  is the same as  found in \cite{Giombi:2014iua}.
To sum over spins we shall follow  the  spectral $\zeta$-function prescription of   \cite{Giombi:2014iua}
which is  equivalent to  introducing the cutoff $e^{-\epsilon\,(s+\frac{d-3}{2})} = e^{-\epsilon\,(s+\frac{3}{2})}$
and dropping all singular terms in the limit $\ep\to 0$, i.e. 
\be\la{43} 
\sum_{s=1}^{\infty} K(s)  \equiv  \sum_{s=1}^{\infty}e^{-\epsilon\,(s+\frac{3}{2})}\,  K(s) \Big|_{\rm \text{finite part}, \ \ep\to 0} 
\ee
Below   we shall   use  the  same prescription also 
for mixed  representations with  $s\equiv \Delta-4$.

One can then  verify the  following relations
\begin{align}
\label{5.3}
& K^{+}(4;\, 0,0,0) +\sum_{s=1}^{\infty} K^{+}(4+s;\, s,0,0) = 0\ , \\
\label{5.4}
& K^{+}(4;\, 0,0,0) +\sum_{s=2,4,\dots }^{\infty} K^{+}(4+s;\, s,0,0) = K_{\phi}\ ,
\end{align}
 where $K_{\phi}=( \aa_\phi, \, E_{c\, \phi})$ are  the real  scalar values  from (\ref{2.2}) and (\ref{2.9}),
As discussed in appendix~\ref{app:singleton}, the l.h.s. of (\ref{5.3})  corresponds to the 
representation content of  the tensor product of two scalar singletons  and 
the associated sum of  characters  is   equal to the  partition function of the singlet sector of the 6d $U(N)$ invariant theory of $N$ 
free complex scalars, see (\ref{C.5}). 
The vanishing to the r.h.s. of \rf{5.4}  is consistent with the expectation  that  the a-anomaly and Casimir energy 
 of the $U(N)$  6d CFT  which are 
proportional to $N$  should  be exactly reproduced  by  the classical  action of  ``non-minimal'' type A   higher spin 
  theory   in \adse   with the inverse coupling $G^{-1}_{\rm non-min} \sim N$,  
so that the one-loop HS correction should vanish  \ci{Giombi:2013fka,Giombi:2014iua}.

The l.h.s. of (\ref{5.4})  corresponds the  field  content of the   ``minimal''  type A   theory in  \adse 
  which should   be dual to  singlet sector of 
 $O(N)$ invariant free real  scalar  6d theory, with the 
 partition function  relation  given by  (\ref{C.8})  (for similar relations in the case of 3d and 4d  cases see \ci{Giombi:2014yra,Beccaria:2014xda}).  Here  the non-vanishing r.h.s.   may be  cancelled  against  part of   the classical  contribution of non-minimal 
  type A theory   
 if one assumed that in this case  $G^{-1}_{\rm min} \sim N-1$   \ci{Giombi:2013fka,Giombi:2014yra}. 
  
Similarly, in  the case when the boundary 6d  theory is    the 
 $U(N)$ invariant free complex (Weyl)  fermion theory 
or $O(N)$ invariant free Majorana-Weyl  fermion theory  (with the dual theory being non-minimal or minimal type B theory in \adse) 
 we get  
\begin{align}
\label{55}
& \sum_{s=1}^{\infty}\Big[K^{+}(4+s;\,s,1,1)+K^{+}(4+s;\,s,0,0) \Big] = 0\ ,\\
&\sum_{s=2,4,\dots}^{\infty}  \, K^{+}(4+s;\,s,1,1)
+ \sum_{s=1,3,\dots}^{\infty}\, K^{+}(4+s;\,s,0,0) = K_{\psi}\ , \la{56}
\end{align}
where the   field content   corresponds to the  one in the  r.h.s.  of  \rf{C.3},(\ref{C.6}) and (\ref{C.9})
and $K_\psi$ is given in \rf{2.2},\rf{2.9}. 
Here we  have  also other representations than  totally symmetric tensors  and thus require general expressions 
in  \rf{3.8},\rf{210}. As in the scalar case, the non-vanishing r.h.s. of \rf{56}  may be   compensated 
 by assuming that the coupling constant of  minimal type B theory is $G^{-1}_{\rm min} \sim N-1$.

When  the 6d boundary theory  is  described   by $N$  real or complex 
 self-dual 2-tensors with dual theory being non-minimal or minimal ``type C''    theory in \adse
 we find (see \rf{C.4},\rf{C.7}, \rf{C.10} and \rf{2.2},\rf{2.9}) 
\begin{align}
\label{57}
& \sum_{s=2}^{\infty} \Big[
K^{+}(4+s;\,s,2,2)+K^{+}(4+s;\,s,1,1)+K^{+}(4+s;\,s,0,0)
\Big] = - K_{T}\ , \\
& \sum_{s=2,4,\dots}^{\infty} \Big[ K^{+}(4+s;\,s,2,2) +K^{+}(4+s;\,s,0,0) \Big] 
+\sum_{s=3,5,\dots}^{\infty}K^{+}(4+s;\,s,1,1)  = \hh\, K_{T}\ . \la{58} 
\end{align}
Here the non-vanishing result is found in both non-minimal and minimal cases. 
This is  similar to what was found in the case of  the AdS$_5$/CFT$_4$  duality 
with the boundary theory  represented by $N$ complex or real Maxwell vectors \ci{Beccaria:2014xda,Beccaria:2014zma}.
The (real) vector corresponds to the parity invariant singleton  combination $\{1\}_{c}=(2;1,0) + (2; 0,1)$ in the $SO(2,4)$ notation.\foot{Here 
we follow \ci{Beccaria:2014xda} and  use the $SU(2) \times SU(2)$ weight  notation for $SO(2,4)$ representation: $(\De; j_1,j_2)$, 
where $h_1= j_1 + j_2, \ h_2  = j_1 - j_2$.}
There  the r.h.s. of the analogs of eqs. \rf{57} and \rf{58}  for the non-minimal and minimal type C theories was the same 
$2K_V$, i.e. twice a  single 4d real vector contribution,  implying the same -2 shift of couplings, i.e.  
$G^{-1}_{\rm non-min} \sim 2N -2$ and $G^{-1}_{\rm min} \sim N -2$.

In the present  case of the  6d self-dual tensor multiplet  theory corresponding to chiral   $\{1\}$   singleton 
  eqs. \rf{57} and \rf{58} imply instead 
$G^{-1}_{\rm non-min} \sim 2N  +1 $ and $G^{-1}_{\rm min} \sim N -\ha $.
Considering instead  the full (self-dual + anti self-dual) tensor 
represented by $\{1\}_{c} = (3;1,1,1) + (3; 1,1,-1) $ (see \rf{C.1},\rf{c12})
one finds that the r.h.s. of the analogs of \rf{57} and \rf{58} become $-2 K_T$ and $0$ respectively
(for  the values of  $E_c$ see \rf{f6},\rf{f10}).  
This  implies that in the \adse  theory dual to the 6d theory  theory of  $N$ complex  6d tensors 
$G^{-1}_{\rm non-min} \sim 2N -1$  and $G^{-1}_{\rm min} \sim 2N$.

The l.h.s.  of the above relations \rf{5.3},\rf{55}  and \rf{57} 
correspond to  $K$ of the  products of singletons $ \{0\} \times \{0\}, \,  \{\hh\}\times  \{\hh\},
$ and $ \{1\} \times \{1\}$  (see  \rf{C.2},\rf{C.3},\rf{C.4}).
One can also  consider a  generalization  when each  factor   in the product   is  a linear combination  of the singletons, i.e. 
$n_{\phi}\, \{0\}+n_{\psi}\, \{\hh\} + n_{T}\,\{1\}$. 
Then (\ref{5.3}),(\ref{55}),(\ref{57})  are generalized to  
\be
\begin{split}
\label{59}
&K^{+}\Big[
\big(n_{\phi}\, \{0\}+n_{\psi}\, \{\hh\} + n_{T}\,\{1\}\big)
\times 
\big(n_{\phi}\, \{0\}+n_{\psi}\, \{\hh\} + n_{T}\,\{1\}\big)
\Big]  \\
 &\qquad \qquad = -n_{T}\,(n_{\phi}\, K_{\phi}+n_{\psi}\, K_{\psi}+n_{T}\, K_{T}),
\end{split}
\ee
where the l.h.s.  is computed   for the representation  content 
appearing in the  character relation in (\ref{C.12}).
For example, in the case when the boundary theory  is decsribed by $N$  complex (2,0) tensor multiplets 
we have $n_\p=5, \ n_\psi=4, \ n_T=1$  we get 
\be 
{K^+ \big( \{\rm tens.\} \times  \{\rm tens.\}\big)} =- K_{\rm tens.} \ , \ \ \ \ \ \qquad 
\{\rm tens.\}  = \{1\}+4 \{\hh\} +5\{0\} \ , \la{510}\ee
where  the tensor multiplet  values of $K_{\rm tens.}$  are given in \rf{33}. 
This  may be compared  with the relation  found in the case of $\N=4$ vector multiplet in 4d  \ci{Beccaria:2014xda,Beccaria:2014zma}:
${K^+ \big( \{\rm vect.  \} \times  \{\rm vect.\}\big)}=2 K_{\rm vect.}$.


\section{One-loop  vacuum energy 
in 10d supergravity on $AdS_{3}\times S^{3}\times M^{4}$}
\label{sec:ads3}

\newcommand{\hhhh}{\frac{3}{2}}

As  discussed  in the Introduction, one  may  also  perform a   similar  one-loop  computations  in the  supergravity sector 
of  type  IIB superstring on  $AdS_{3}\times S^{3}\times T^{4}$ to determine the subleading  term in the central charge \rf{19} 
or the vacuum energy \rf{20}.\foot{String modes corresponding to massive unprotected multiplets  are expected not to contribute to $c$.}

The one-loop $AdS_3$  vacuum energy can be computed by starting with 
the  spectrum of 6d supergravity  on  $AdS_{3}\times S^{3}$
as  massive KK  multiplets  on $M^{4}=T^4$  should  not contribute  due to supersymmetric cancellation.
More generally,   we may consider in parallel  the cases of IIA or IIB supergravities on 
$M^4=T^4$ or $K3$. The results for the one-loop vacuum energy are expected to be the  same.\foot{For example, 
type IIB  theory  on  $AdS_{3}\times S^{3}\times T^{4}$   with RR 3-form flux  is S-dual to type IIB  theory with NSNS    flux 
and as the supergravity theory is S-duality invariant  the same should be true  for  the value of $E_c$. Since NS-NS sector is common 
to IIB and IIA   theories, the same  result should be found also  in the corresponding  IIA theory.} 

The list of relevant 6d supergravities with $\N = (n_{L}, n_{R})$ supersymmetry  was given in 
\cite{deBoer:1998ip}, where an algorithm  for construction  of the corresponding 
 KK spectrum on $S^3$ was   presented. Below  we shall consider the following cases:
\be
\la{5a}
\begin{array}{|c|c|c|}
\hline
\text{10d} & M^{4} & (n_{L}, n_{R}) \\
\hline
\text{IIB} & K3 & (2,0) \\
\text{IIA} & K3 & (1,1) \\
\text{IIA or IIB} & T^{4} & (2,2)\\
\hline
\end{array}
\ee

\

\subsection{KK towers of states on $S^3$ }  

The  6d supergravity  fields  transform 
in representations $(j_{1},j_{2})$ of the 6d little group $SO'(4)\simeq SU(2)\times SU(2) $  (of $SO(1,5)$ in the tangent space). 
This gives a set $\Phi$ of 
representations of the diagonal    subgroup $SO(3)\simeq SU(2)$ of $SO(4)$. 
Considering compactification on 
  $S^3$, the above $SO(3)$   can be  identified  with the  factor in  $S^3=SO(4)/SO(3)$. 
Each representation $R\in \Phi$ is associated with a tower of KK states with $SO(4)$ representations
containing $R$ under restriction to their diagonal $SO(3)$. 

These KK  fields carry also representation  of the $AdS_3$  isometry group $SO(2,2)$   (or global part of 2d conformal group)  which are 
are labelled by scaling dimension and spin $(\Delta, s)$, with 
$\Delta \ge |s|$. The values of $(\Delta, s)$ can be determined by re-organizing the KK towers in 
short supermultiplets of $SU(2,2\,|1)\times SU(2,2\,|1)$ since its generators include the dilatation (Virasoro $L_{0}$)
and spin operators.  The relevant short representations $(J)_{\rm s}$ of $SU(2,2\,|1)$ have the following  content  
\be
\label{5.2}
(J)_{\rm s}: \qquad 
\begin{array}{c|cc}
\text{states} & j & L_{0}  \\
\hline
|0\rangle & J & J \\
Q_{\pm} |0\rangle & J-\hh & J+\hh  \\
Q_{+}Q_{-} |0\rangle & J-1 & J+1  
\end{array}
\ee
where $|0\rangle$ is the lowest weight of the representation in the usual oscillator construction \cite{Gunaydin:1986fe}, 
$Q_{\pm}$ are the supercharges, and $j$ is  $SU(2)$ spin. 
Thus, in  general, each short $(J)_{\rm s}$ representation contains four $SO(2,2)$ representations. 
Using  (\ref{5.2})  and that   $\Delta=L_{0}+\overline L_{0}$, \ \  $s = L_{0}-\overline L_{0}$    one obtains 
the  quantum numbers  of representations in  the 
tensor products $(\overline J, J)_{\rm s}$.

Let us now  list the KK towers that  appear  in the  theories in \rf{5a}. 
 For $(2,0)$  6d supergravity, or IIB  theory dimensionally reduced on K3 
the field content is a graviton, five self-dual two-forms, four gravitinos, and $n_{T}=21$ tensor multiplet 
of one anti self-dual two-form, four fermions and five scalars (see  also   \cite{Deger:1998nm,Nicolai:2003ux})\foot{We  shall  keep $n_{T}$ generic because this will be useful in comparing 
with IIA case.} 
\be
\label{5.33}
\begin{split}
\Phi^{(2,0)} &= 
(1,1)+4\,(\hh, 1)+5\,(0,1)+n_{T}\,\big[
(1,0)+4\,(\hh,0)+5\,(0,0)
\big], \\
& \  \ \   \ \ \ g_{\mu\nu} \ \ \ \ \  \ \ \ \  \psi_{\mu} \ \ \ \  \ \ \ \ B 
\qquad\qquad\quad \widetilde B \qquad \quad\psi \qquad\quad\ \varphi
\end{split}
\ee
Reorganising  KK towers in short multiplets of 
$SU(2,2\,|1)\times SU(2,2\,|1)$, we find 
\be
\label{5.44}
\begin{split}
\Phi_{\rm KK}^{(2,0)} &= \sum_{\ell=0}^{\infty} \Phi_{2}(\ell)+(n_{T}+1)\,
\sum_{\ell=0}^{\infty} \Phi_{1}(\ell)+n_{T}\,(\hh,\hh)_{\rm s}\ , 
\end{split}
\ee
where
\be
\label{5.5}
\begin{split}
\Phi_{2}(\ell) &= \te \big(\frac{\ell+1}{2},\frac{\ell+3}{2}\big)_{\rm s} + \big(\frac{\ell+3}{2},\frac{\ell+1}{2}\big)_{\rm s}\ , \qquad 
\qquad
\Phi_{1}(\ell) =\te \big(\frac{\ell+2}{2},\frac{\ell+2}{2}\big)_{\rm s}.
\end{split}
\ee
The towers in the first and second sums are called spin-2 and spin-1 towers because of the maximum spin 
of their bottom floor $\ell=0$. The explicit field content is collected in Appendix ~\ref{app:tower} and their 
 6d origin  is discussed  in \cite{Deger:1998nm}.
 
For $(1,1)$ 6d supergravity, or 10d IIA  supergravity   reduced on K3, the field content is the sum of  6d graviton multiplet  and 
 $n_{V} = 20$  vector multiplets \cite{Strathdee:1986jr}. The $SO(4)$ little group representations are\footnote{Here  we combine 
representations related by conjugation $(j_{1},j_{2})\to (j_{2},j_{1})$ since they  give same contribution to KK spectrum.}
\be
\label{5.6}
\begin{split}
\Phi^{(1,1)} &= 
(1,1)+4\,(\hh, 1)+2\,(0,1)+4\,(\hh,\hh)+4\,(\hh,0)+\,(0,0)\\
&\quad + n_{V}\,\Big[(\hh,\hh)+4\,(\hh,0)+4\,(0,0)\Big],
\end{split}
\ee
and the KK towers are 
\be
\label{5.7}
\begin{split}
\Phi_{\rm KK}^{(1,1)} &= \sum_{\ell=0}^{\infty} \Phi_{2}(\ell)+(n_{V}+2)\,
\sum_{\ell=0}^{\infty} \Phi_{1}(\ell)+(n_{V}+1)\,(\hh,\hh)_{\rm s}.
\end{split}
\ee
Comparing (\ref{5.44}) and (\ref{5.7}), we see that they are equal under the identification 
$n_{V}+1 = n_{T}$ that is indeed true for the physical values. Thus we  should find that $
E_{c}({\text{IIB on K3}}) = E_{c}({\text{IIA on K3}}) $
(as was already   mentioned above,   this is  implied by S-duality of IIB theory and NS-NS sector  being common for IIA and IIB theories). 

Finally, for $(2,2)$ 6d supergravity, or IIA or IIB theory  dimensionally reduced  on $T^{4}$ the field content is 
a graviton, five self-dual and five anti self-dual two-forms, eight gravitinos, 16 gauge fields, 40 fermions and 25 scalars:
\be
\label{5.9}
\begin{split}
\Phi^{(2,2)} &= 
(1,1)+8\,(\hh, 1)+5\,(0,1)+5\,(1,0)+16\,(\hh,\hh)+40\,(\hh,0)+25\,(0,0)\   \\
& \ \ \ \  \ \ g_{\mu\nu} \ \ \ \ \ \ \ \ \  \psi_{\mu} \ \ \ \ \ \  \ \ B 
\qquad\quad \widetilde B \qquad \qquad V_{\mu} \qquad\quad\ \ \ \psi\quad\qquad \ \ \ \varphi
\end{split}
\ee
The  KK towers  here are 
\be
\label{5.10}
\begin{split}
\Phi_{\rm KK}^{(2,2)} &= \sum_{\ell=0}^{\infty} \Phi_{2}(\ell)
+4\,\sum_{\ell=0}^{\infty}\Phi_{3\ov 2 }(\ell)+6\,\sum_{\ell=0}^{\infty} \Phi_{1}(\ell)
+5\,(\hh,\hh)_{\rm s}\ , 
\end{split}
\ee
where
\be
\label{5.11}
\Phi_{3\ov 2}(\ell) =\te  \big(\frac{\ell+1}{2},\frac{\ell+2}{2}\big)_{\rm s}  +  \big(\frac{\ell+2}{2},\frac{\ell+1}{2}\big)_{\rm s} 
\ee
is 
a fermionic spin-$3 \ov 2 $ tower  (see Appendix ~\ref{app:tower}).

\subsection{ Vacuum  energy }   

The $AdS_3$   vacuum energy  contributions  of the above KK towers 
  can be computed using the expressions for the characters 
or one-particle  partition functions of the corresponding $SO(2,2)$  representations  which we shall first  recall. 

$SO(2,2)$  viewed as  global conformal group in 2d   is generated by the  $L_{0}, L_{\pm 1}$
and $\overline L_{0}, \overline L_{\pm 1}$ Virasoro generators. 
Unitary irreducible representations of $SO(2,2)$ are massive for $\Delta > |s|$ and massless for $\Delta=|s|$.
A massive representation is built on a ground state $|h,\overline h\rangle$ with $h\overline h>0$. Thus, both $L_{-1}$
and $\overline L_{-1}$ give a non zero result  and the resulting character  is (see, e.g.,  \cite{Gunaydin:1986fe})\foot{The double factor of $1/(1-q)$ takes into account multiple applications of both $L_{-1}$ and $\overline L_{-1}$.}
\be
\label{5.12}
\Delta > |s|: \qquad\qquad 
 \widehat\Z^{+}(\Delta; s) = \frac{q^{\Delta}}{(1-q)^{2}},
\ee
A massless representation with $\Delta=|s|>0$ has conformal weights 
$(h,0)$ or $(0,\overline h)$. Acting with the lowering operators $L_{-1}$ and $\overline L_{-1}$ 
on $|h,\overline h\rangle$
only one of them gives a non-zero result. As a consequence, here 
\be
\label{5.13}
\Delta = |s|: \qquad \Z^{+}(|s|; s) = \frac{q^{\Delta}}{1-q} = \frac{q^{\Delta}-q^{\Delta+1}}{(1-q)^{2}} = 
\widehat\Z^{+}(\Delta; s)-\widehat\Z^{+}(\Delta+1; s),
\ee
Finally, for $\Delta=s=0$, 
we have only the ground state $|0,0\rangle$ and $\Z^{+}(0; 0) = 1$.  The expressions (\ref{5.12}) and 
(\ref{5.13}) can be used to prove that $SU(1,1\,|\,2)$ short multiplets obey the important relation 
$E_{c} = -\frac{1}{12}\,c$, see (\ref{20}). We discuss this in details in Appendix \ref{app:cardy}.


The   contribution from a particular $SO(2,2)$   representation to the 
 $AdS_3$ vacuum   or $S^1$ 2d  Casimir energy $E_c$ can then  be computed using \rf{2.7},\rf{2.8}. 
Explicitly, for a massive field in $AdS_{3}$,
we may   write the partition function (\ref{5.12}) as 
\be\la{5.14}
\widehat  \Z^{+}(\Delta; s) = \sum_{n=0}^{\infty}(n+1)\,q^{\Delta+n}.
\ee
We then  obtain a formal (divergent) expression for the corresponding  $E_{c}$  as (cf. \rf{37},\rf{4.7})
\ba
\label{5.15}
\widehat E^+_{c}(\Delta;\; s) = \sum_{n=0}^{\infty}  e_n(\Delta;\; s), \qquad \qquad 
e_n (\Delta;\; s) =
\frac{1}{2}\,(-1)^{2\,s} \,(n+1)\,(\Delta+n)\  .
\end{align}
In addition, we then need to sum over the KK states. 

There will be divergences  coming from the sum over $n$, but also from the sum over the KK level $\ell$.
Like in  $AdS_5 \times S^5$ case   \ci{Beccaria:2014xda}   and $AdS_7 \times S^4$ case in section 3
the total   sum  may  be  again computed   using  the  $\zeta$-function  regularization applied   to
the full effective 6d energy  eigenvalue $\Delta+n$, or, equivalently, by  introducing 
the cutoff  
$
e_n \   \to \ \ e_n \, e^{- \epsilon (\Delta +n)} \ ,  
$
doing  the sum, expanding  in $\ep \to 0$, and  dropping  all singular terms. 
Applying this procedure 
to  theKK towers  appearing  in \rf{5.44},\rf{5.7} and \rf{5.10}, we obtain 
\ba
\label{5.16}
 E_{c, 2} = E_{c}\big[\sum_{\ell=0}^{\infty}\Phi_2(\ell) \big] = \te  -\frac{89}{192},  & \te 
\ \qquad  E_{c, {3 \ov 2} } = E_{c}\big[\sum_{\ell=0}^{\infty}\Phi_{3\ov 2 }(\ell)\big ] = \frac{19}{96}, \\
E_{c, 1} = E_{c}\big[\sum_{\ell=0}^{\infty}\Phi_1(\ell) \big] =\te  -\frac{101}{384}, &
\ \qquad  E_{c, \rm extra} = E_{c}\big[(\hh,\hh)_{\rm s} \big] = \te  \frac{1}{4}\ , \la{517}
\end{align} 
where $E_{c, \rm extra}$ is the contribution from the $(\hh,\hh)_{\rm s}$ representation appearing in 
(\ref{5.44}), (\ref{5.7}), and (\ref{5.10}) in the bottom part of the KK towers. 

The  above  are the contributions from the 
massive $SO(2,2)$ representations. 
As discussed in \cite{Deger:1998nm,deBoer:1998ip}, the resolution of the missing states puzzle raised 
in \cite{Vafa:1998nt} amounts to the re-introduction of the massless representations ($\ell=-1$ states in 
the spin 2 and $\hhhh$ towers). These are massless  multiplets in $AdS_{3}$ that do not carry propagating degrees of freedom. 
Their structure is presented in Appendix ~\ref{app:tower}. For these multiplets we find 
\be
\label{5.17}
\te E_{c, 2}^{\rm massless} = E_{c}\big[(0,1)_{\rm s}+(1,0)_{\rm s}\big] = \frac{1}{2},\qquad\qquad 
E_{c, \hhhh}^{\rm massless} = E_{c}\big[(0,\hh)_{\rm s}+(\hh,0)_{\rm s}\big] = -\frac{1}{4}.
\ee
Collecting all  contributions of states in  (\ref{5.44}), we find in the case of  for IIB theory on K3   
\be
\label{5.18}
\begin{split}
E_{c}^{(2,0)} &= E_{c, 2}^{\rm massless}+E_{c,2}+(n_{T}+1)\,E_{c,1}+n_{T}\,E_{c, \rm extra}\\
&\te = \frac{1}{2}-\frac{89}{192}-(n_{T}+1)\frac{101}{384}+n_{T}\,\frac{1}{4} = -\frac{29}{128}-\frac{5}{384}\,n_{T}\ \  \stackrel{n_{T}=21}{\longrightarrow}\  -\frac{1}{2} \ . 
\end{split}
\ee
This is also the result for IIA theory on K3, as follows from (\ref{5.7}). 
From   (\ref{5.10})
we also get  exactly the same   result   for  IIA or IIB  theory on $T^{4}$,
\be
\label{5.19}
\begin{split}
E_{c}^{(2,2)} &= E_{c, 2}^{\rm massless}+E_{c,2}+4\,\big(E_{c,\hhhh}^{\rm massless}+E_{c, \hhhh}\big)
+6\, E_{c,1}+5\,E_{c, \rm extra}\\
&\te = \frac{1}{2}-\frac{89}{192}+4\,\big(-\frac{1}{4}+\frac{19}{96}\big)-6\,\frac{101}{384}+5\,\frac{1}{4}
= -\frac{1}{2} \ , 
\end{split}
\ee
in agreement with the claim in \rf{20}. 


\section*{Acknowledgments}
We  thank R. Roiban, K. Zarembo, and R. Metsaev  for discussions. 
A.A.T.  acknowledges the  hospitality and support of the 
  Simons Center for Geometry and Physics, Stony Brook University where   part of this work was done. 
The  work of A.A.T is supported by the ERC Advanced grant No.290456. 
The work of  M.B.  and  A.A.T.  is  part of collaboration supported 
 by  the  Russian Science Foundation grant 14-42-00047 and associated with Lebedev Institute.

\appendix

\def \rmR {{\mathsf R}}  
\def \rmH {{\mathsf H}}
\def \rmG {{\mathsf G}}
\def \ru {\ell}

\section{Free  (2,0)  multiplet in 6d}
The field content of the $(2,0)$ tensor multiplet is composed of five scalars fields $\phi^{a}$, 
two complex Weyl fermions $\psi^{I}_{L}$ or 4  Majorana-Weyl fermions (each with 4 real components), 
and an antisymmetric tensor $T_{ij}$ with (anti) selfdual strength.
It  reprsents a free  6d  CFT invariant  under  superconformal $\mathcal{N}=(2,0)$   group  \cite{Nahm:1977tg, Howe:1983fr} containing  the  conformal group  $SO(2,6)$  and 
the  R-symmetry  group   $SO(5)\simeq USp(4)$. 

The Weyl anomaly of the 
(2,0) multiplet  was  discussed in \cite{Bastianelli:2000hi}.
The values of the a-anomaly coefficients  for  the individual   fields are
(here ${\psi}$  stands  for   one 6d Majorana-Weyl fermion) 
\be
\label{2.2}\te 
\aa_{\phi} = -\frac{1}{72576},\qquad\ \ \ 
\aa_{\psi} = -\frac{191}{1451520},\qquad\ \ \ 
\aa_{T} = -\frac{221}{40320}.
\ee
The total a-anomaly of one free $(2,0)$ tensor multiplet  is thus
\be\la{a2}\te 
\aa_{\rm tens.} = 5\,\aa_{\phi}+4\,\aa_{\psi} +\aa_{T} =  -\frac{7}{1152}\ .
\ee
Considering (2,0)  multiplet on $S^1\times S^5$  one may  compute the corresponding thermal 
partition function. 
The canonical (or one-particle) partition function of a free CFT in $S^{1}\times S^{d-1}$      
can be computed by direct evaluation of the free QFT path-integral in terms of the eigenmodes of the
 quadratic kinetic operator. 
An alternative approach is the 
 operator counting method \cite{Cardy:1991kr,Kutasov:2000td,Beccaria:2014jxa}. 
From the spectrum  of    eigenvalues of the Hamiltonian or dilatation operator 
$\omega_{n} = \Delta_{n}$ and their
degeneracies ${\rm d}_n$   one    gets 
\be
\label{2.4}
\Z(q) = {\rm Tr}\, e^{-\beta H} =  \sum_n  {\rm d}_n\, e^{-\beta\,\omega_n} =   \sum_n  {\rm d}_n\,   q^{\Delta_n},
\qquad q\equiv e^{-\beta}\ .
\ee
In the  approach based on  counting of states 
one  needs to consider the contribution of 
 off-shell components (and  their  derivative descendants)  of a suitable gauge invariant 
  field strength  modulo non-trivial gauge identities  and then subtract  
  the components  of  the  equations of motion for the field strength (and their  
  derivatives). 
 The single particle partition functions for the 5 scalars, 
4 Majorana-Weyl fermions, and self-dual tensor in $S^{1}\times S^{5}$   are 
 \cite{Kutasov:2000td}  
\be\la{a4}
\begin{split}
\Z_{\phi}(q) &= \frac{1}{12}\,\sum_{n=0}^{\infty}(n+1)(n+2)^{2}(n+3)\,q^{n+2} = 
\frac{q^{2}-q^{4}}{(1-q)^{6}}\ ,\\
\Z_{\psi}(q) &= \frac{1}{6}\,\sum_{n=0}^{\infty}(n+1)(n+2)(n+3)(n+4)\,q^{n+\frac{5}{2}} = 
\frac{4\,q^{\frac{5}{2}}-4\,q^{\frac{7}{2}}}{(1-q)^{6}}\ ,\\
\Z_{T}(q) &= \frac{1}{4}\,\sum_{n=0}^{\infty}(n+1)(n+2)(n+4)(n+5)\,q^{n+3} = 
\frac{10q^{3}-15q^{4}+6q^{5}-q^{6}}{(1-q)^{6}}\ .
\end{split}
\ee
These expressions are in agreement with \rf{26},\rf{25},\rf{3.2}. 

The related Casimir energy on $S^{5}$ can be computed from the one-particle partition function $\Z(q)$ 
using \rf{2.7},\rf{2.8}:
\be
\label{2.9}\te 
E_{c, \phi} = -\frac{31}{60480}, \qquad \ \ 
E_{c, \psi} = -\frac{367}{96768}, \qquad \ \ 
E_{c, T} = -\frac{191}{4032}.
\ee
Then the  total Casimir energy for  the free (2,0) tensor multiplet is
\be
\label{2.10}\te 
E_{c,\,\rm tens.} = 5\,E_{c, \phi}+4\,E_{c, \psi} +E_{c, T} = -\frac{25}{384}\ . 
\ee
 This  agrees with the   value  found   in  \cite{Gibbons:2006ij}.\footnote{Note that 
the ratio of the vacuum energy \rf{2.10}  and the a-anomaly \rf{a2}, i.e.  
  $E_{c,\, \rm tens.}/\aa_{\rm tens.} = \frac{75}{7}$, 
differs from  the  expression  in   \cite{Herzog:2013ed}. 
The reason is that  the Casimir energy is  computed in the standard 
$\z$-function regularization scheme in which derivative terms $D_6$ 
 in the conformal anomaly \rf{1} do not vanish  \cite{Bastianelli:2000hi}  while ref. \cite{Herzog:2013ed}
 assumed an abstract scheme where  there are no derivative terms in the anomaly 
 (see also a related discussion in 
\cite{Beccaria:2014xda}). }




Let us note  that the expressions in \rf{a4} admit also \adse  interpretation. 
In general,  given a conformal  6d field, the corresponding  
  one-particle partition function $\Z(q)$ may be  expressed as    \ci{Beccaria:2014jxa}
\be \la{a7} 
\Z(q) = \Z^{-}(q)-\Z^{+}(q)\ , 
\ee
where $\Z^{\pm}(q)$ are the  one-particle  partition functions   
for  the one-loop  partition  function $Z^\pm$  of the  associated     higher spin  field
in (thermal quotient of) \adse    computed with the 
standard  (``Dirichlet'')  or alternative (``Neumann'')    boundary conditions. 
The canonical dimension of the  conformal 6d field is equal to $\Delta_- = 6 - \Delta, \ \Delta=\Delta_+$. 
For generic representation  \rf{a7}   may be written as 
\be \la{a9}
\Z(q) = \Z^{+}(\Delta;\, \mathbf{h})( q^{-1})-\Z^{+}(\Delta;\, \mathbf{h})( q)+\sigma(q) \ , 
\ee
where  $\sigma(q)$ can be interpreted as a Killing tensor character 
 associated with missing gauge 
invariances \cite{Beccaria:2014jxa}. This term is a polynomial in $q$ and $1/q$   which is  symmetric under $q\to 1/q$
For a 6d  conformal  scalar  with canonical dimension 2 we find that (see \rf{3.2},\rf{25},\rf{26},\rf{a4}) 
\be
\label{B.4}  \Z_{\phi}(q)=
\Z^{+}(4;\, 0,0,0)( q^{-1})-\Z^{+}(4;\, 0,0,0)( q) \ . 
\ee 
For the Majorana-Weyl 6d fermion with canonical dimension  $\textstyle\frac{5}{2}$ we get 
\be
\label{B.5} \Z_{\psi}(q) = 
\Z^{+}({\textstyle\frac{7}{2}};\, \hh, \hh, \hh)(q^{-1})
-\Z^{+}({\textstyle\frac{7}{2}};\, \hh, \hh, \hh)( q) \ . 
\ee 
In the case of   rank 2 tensor  of dimension 2    let  us  define  (see \rf{3.2},\rf{255})
\be
\label{B.6}
\Z^{+}_{T} \equiv  \Z^{+}(4;\, 1,1,0) =  \widehat\Z^{+}(4;\, 1,1,0)-\widehat\Z^{+}(5;\, 1,0,0)+\widehat\Z^{+}(6;\, 0,0,0)\ .
\ee
This is a $SO(2,6)$ character corresponding to the   massless  case (unitarity bound)
in (ii) in \rf{21}.\foot{Here the two additional terms   are related to gauge freedom 
 in the rank 2 tensor  potential.}
One observes then  that 
\be
\label{B.7}  2\,\Z_{T}(q) = 
\Z^{+}_{T}(q^{-1})-\Z^{+}_{T}(q) -1 \ ,
\ee
where $\Z_T$   is the self-dual tensor partition function in \rf{26},\rf{a4}. 
The -1 term   should be  interpreted as  a subtraction of a non-normalizable gauge transformation.

Note that in  the case  of the (2,0) tensor multiplet corresponding  formally to the $p=1$  singleton level of KK tower in 
 Table 1   we find 
\be
\label{D.3}
\begin{split}
\Z_{\rm tens.}(q) =  5\,\Z_{\phi}+4\,\Z_{\psi}+\Z_{T} = \frac{
5\,q^{2}+16\,q^{\frac{5}{2}}+15\,q^{3}-5\,q^{4}+q^{5}
}{(1-q)^{5}} \ , 
\end{split}
\ee
which satisfies the relation 
\be
\label{D.4}
\Z_{\rm tens.}(q)+\Z_{\rm tens.}(q^{-1})+1=0\ .
\ee
 The  general  relations for the boundary  conformal anomaly and Casimir energy are  \cite{Beccaria:2014xda}
\be\la{a8}
\aa = -2\aa^{+}, \qquad\qquad  E_{c} = -2\, E_{c}^{+} \ .
\ee
Denoting  $K=(E_{c}, \aa)$ and $K^{+}=(E_{c}^{+}, \aa^{+})$, we have\footnote{In the 
tensor case, the  factor $\hh$   is absent   due to  the self-duality condition.}
\be
K^{+}(4;\, 0,0,0) =  -\hh \,K_{\phi},  \qquad
K^{+}({\textstyle\frac{7}{2}};\, \hh, \hh, \hh) =  -\hh \,K_{\psi}, \qquad
 K^{+}(4;\, 1,1,0)  =  -\,K_{T},  
\ee
where  $K_\phi,K_\psi,K_T$ are given by  (\ref{2.2}) and (\ref{2.9}).

\section{a-anomaly from spectral $\zeta$-function in $AdS_7$}
\label{app:zeta}

The a-coefficient of the 
boundary conformal anomaly   can be determined from 
 the  logarithmic IR   singular part of the 
 one-loop partition function 
 in Euclidean \adse    with boundary $S^6$, i.e.   hyperboloid  $\mathbb H^{7}$
 (see, e.g., \ci{Diaz:2007an,Giombi:2013yva})
 \be\la{a1}
 \log Z^{+} = -\ha\log{\det}_{+}\, \mc O = \ha\,\zeta'(0) = - 96\, \aa^+ \log \RR  + ... \ .
 \ee
 Here  $\zeta(z)$ is the spectral zeta function found  by  evaluating the trace of the  $\mathbb H^{7}$ 
 heat kernel \cite{Camporesi:1994ga} 
 associated with the 7d operator $\mc O$  and $\RR$  is an   IR cutoff
  regularising  the  volume of $\mathbb H^{7}$.
  
 Below  we shall consider  the  operator $\mc O$   corresponding to a generic massive (or massless)  higher spin field   in    representation $(\Delta;\, \mathbf{h})$ 
  generalizing the expression in \ci{Giombi:2013yva}  found in the totally symmetric tensor case
  $h_1=s, \, h_2=h_3=0$ \foot{For the general  form of $X$ see  
 \cite{Metsaev:1994ys,Metsaev:1995re,Metsaev:2003cu}.}
\be\la{a22}
  \OO =-D^{2}+X\,, \qquad \qquad  X =  \Delta\, (\Delta-6)  - h_{1} -h_{2}- |h_{3}| \ . 
 \ee
Here $D^{2}$ is the standard Laplacian in $AdS_7$ defined on transverse  field. 
  The  discussion will  be parallel to the one in $AdS_5$ case in \ci{Beccaria:2014xda}.
 
The spectral $\zeta$-function of the operator $\OO$ can be expressed in terms of the heat kernel 
\be
\zeta(z) =  \frac{1}{\Gamma(z)}\int_{0}^{\infty}dt\, t^{z-1}\,{\rm Tr }\, K \ , \ \ \ \ \qquad 
K(x, y; t) = \langle x|e^{-t\,\mc O}|y\rangle  \ .  
\ee
Since  $\mathbb H^{7}$ is homogeneous,  the trace  over the  position $x$  
gives a factor of (regularized) volume, i.e. 
\be
\zeta(z) = \mbox{Vol}(\mathbb H^{7})\,\zeta(z; x) \ , \ \ \ \ \ \ 
 \zeta(z; x) \equiv   \frac{1}{\Gamma(z)}\int_{0}^{\infty}dt\, t^{z-1}\,{\rm tr }\, K(x,x; t),
\ee
where $\tr$ is the  trace over the  representation indices of the operator  and $\zeta(z; x)$ does not actually depend  on $x$. 

One  can use the results  for  the heat kernel 
of the Laplacian  in $AdS_{2n +1}$   with  even  $n$   derived in  
\cite{Camporesi:1994ga,Gopakumar:2011qs} 
applying them to the case  of $n=3$. 
It is convenient to  
 start  with   heat-kernel for  the sphere $S^{7}$ and then  analytically continue to  $AdS_{7}$. 
Let us consider a field on $S^{7}$ transforming under the tangent space rotations  in a  
representation $\rmG$ of $SO(7)$.
Since S$^{7}=SO(8)/SO(7)$, 
 the heat kernel receives contributions from each representation 
$\rmR$ of $SO(8)$ that contains $\rmG$ when restricted to $SO(7)$.
 Let us  denote  $\rmR$ and $\rmG$  by  the corresponding weights as 
\be
\begin{split}
\rmR = (\ru_{1}, \ru_{2}, \ru_{3}, \ru_{4}), \quad \ru_{1}\ge \ru_{2}\ge \ru_{3} \ge |\ru_{4}|, \quad 
\rmG = (g_{1}, g_{2}, g_{3}),  \quad \ \ g_{1}\ge g_{2} \ge g_{3} \ge 0,
\end{split}
\ee
were all labels are integer or half integer. The  branching condition on the representation 
$\rmR$ is  \be \la{A.6} 
\ru_{1}\ge g_{1} \ge \ru_{2}\ge g_{2} \ge \ru_{3} \ge g_{3} \ge |\ru_{4}|,
\ee 
with the additional requirement that $\ru_{i}-g_{i}\in \mathbb Z$. 
The heat kernel at the coincident points, traced over
representation indices, can be written  as 
\be
{\rm tr}\,  K(x,x; t) = \frac{3}{\pi^{4}}\,\sum_{\ru_{i}}\, \dd_{\rm R}\, e^{-t\,E_{\rmR}^{(\rmG)}},  
\ee
where  $E_{\rmR}^{(\rmH)}$  are  the eigenvalues of the Laplacian $ -D^2$  on  S$^7$
expressed in terms of the second  Casimir   values for the two representations  and  $d_{\rmR}$ is the dimension of $\rmR$
\ba
& \left. -  D^{2}\right|_{\rm S^{7}}\  \to \    E_{\rm R}^{(\rmG)} =  C_{2}({\rmR})  -  C_{2}({\rmG}) \  , \\
& C_{2}({\rmR}) = \ru_4^2+\ru_1 \left(\ru_1+6\right)+\ru_2 \left(\ru_2+4\right)+\ru_3 \left(\ru_3+2\right), \\
& C_{2}({\rmG}) =g_3^2+g_3+g_1 \left(g_1+5\right)+g_2 \left(g_2+3\right), \\
&\dd_{\rmR} = \frac{1}{4320}\left[\left(\ru_1+3\right){}^2-\left(\ru_2+2\right){}^2\right]  \left[\left(\ru_1+3\right){}^2-\left(\ru_3+1\right){}^2\right]\no \\
&\qquad  \times \label{A.11} \left[\left(\ru_2+2\right){}^2-\left(\ru_3+1\right){}^2\right]\left[\left(\ru_1+3\right){}^2-\ru_4^2\right]
 \left[\left(\ru_2+2\right){}^2-\ru_4^2\right]
   \left[\left(\ru_3+1\right){}^2-\ru_4^2\right]\ .  
\end{align}
The analytic continuation from $S^7$ to $AdS_7$  amounts to \cite{Camporesi:1994ga,Gopakumar:2011qs}
\be
\label{A.12}
\ru_{1}\to i\,\lambda-3\ ,
\ee
with the sum over $\ru_{1}$ becoming an integral over $\lambda\ge 0$. Finally, considering 
 states saturating the 
inequalities (\ref{A.6}) and identifying $(\ru_{2},\ru_{3},\ru_{4}) = \mathbf{h}=(h_1,h_2,h_3)$,
we find  that the eigenvalues of the operator \rf{a22} for the representation $(\Delta;\, \mathbf{h})$  are 
\be
\left. (-D^{2}+X)\right|_{\rm AdS_{7}}  \ \to \ \  \lambda^{2}+(\Delta-3)^{2} \ .  
\ee
The  regularised volume  
 may be written as 
$\mbox{Vol}(\mathbb H^{7}) = \textstyle\frac{1}{3}\pi^{3}\, \log \RR +...$ where  the IR  cutoff 
$\RR$ is  the radius of $S^6$ measured in 7d metric  $d\rho^{2} + \sinh^{2}\rho\, d\Omega^{2}_6$
at  large $\rho$.
Doing the analytic continuation (\ref{A.12}) in  the dimension $\dd_{\rmR}$ in \rf{A.11}
we  finally obtain 
\be\la{a14}
\begin{split}
\zeta(z) &=  \text{Vol}(\mathbb{H}^{7})\, \zeta(z; x) \\
& \to -\frac{\log \RR}{4320 \pi} \left(h_1-h_2+1\right) \left(h_1+h_2+3\right) \left[\left(h_1+2\right){}^2-h_3^2\right)\left(\left(h_2+1\right){}^2-h_3^2\right]\\
&
\times  \int_{0}^{\infty} d\lambda\,  \frac{ \left[h_1
   \left(h_1+4\right)+\lambda ^2+4\right] \left[h_2 \left(h_2+2\right)+\lambda ^2+1\right] \left(h_3^2+\lambda ^2\right)}{\left[\lambda^{2}+(\Delta -3)^2\right]^{z}} \ . 
\end{split}
\ee
Integrating over  $\lambda$ and taking the $z$-derivative  at $z=0$
 we may  then use \rf{a1} to  find 
the expression for $\hat \aa^+$  in  (\ref{210}).

\section{Tensor products of $SO(2,6)$ singleton representations  and associated character relations}
\label{app:singleton}

Let us introduce the following notation for the  spin $j=0, \hh, 1, ...$ singleton representations of $SO(2,6)$
\be
\label{C.1}
\begin{split}
\{j\} = (2+j;\, j,j,j)\ .
\end{split}
\ee
Here $\{0\} $ corresponds to a real  scalar $\p$, $\{\hh\} $ to  MW fermion $\psi$, and $ \{1\}$ to self-dual  tensor $T$. 
We shall also use the  notation $(\Delta; h_1,h_2,h_3)_{ c} \equiv  (\Delta;\, h_1,h_2,h_3  )+(\Delta; h_1,h_2,-h_3)$, so that 
$\{j\}_c = (2+j;\, j,j,j)+  (2+j;\, j,j,-j)$.

From the general  Flato-Fronsdal relations  in  \cite{Vasiliev:2004cm,Dolan:2005wy} we get  in the  present 6d case 
\begin{align}
\label{C.2}
\{0\}\times \{0\} &= (4;\, 0,0,0)+\bigoplus_{s=1}^{\infty}(4+s;\, s,0,0)\ ,\\
\{\hh\}\times \{\hh\} &= \bigoplus_{s=1}^{\infty}\Big[
(4+s;\,s,1,1)+(4+s;\,s,0,0) 
\Big]\ , \label{C.3} \\
\label{C.4}
\{1\}\times \{1\} &= \bigoplus_{s=2}^{\infty} \Big[
(4+s;\,s,2,2)+(4+s;\,s,1,1)+ (4+s;\,s,0,0)
\Big]\ .
\end{align}
The  above  relations imply analogous relations for the characters or one-particle partition functions
\begin{align}
\label{C.5}
\big[\Z_{\phi}(q)\big]^{2} &= \Z^{+}(4;\,0,0,0)+\sum_{s=1}^{\infty}\Z^{+}(4+s;\, s,0,0),\\
\big[\Z_{\psi}(q)\big]^{2} &= \sum_{s=1}^{\infty}\Big[
\Z^{+}(4+s;\,s,1,1)+\Z^{+}(4+s;\,s,0,0) 
\Big], \label{C.6} \\
\label{C.7}
\big[\Z_{T}(q)\big]^{2} &= \sum_{s=2}^{\infty} \Big[
\Z^{+}(4+s;\,s,2,2)+\Z^{+}(4+s;\,s,1,1)+\Z^{+}(4+s;\,s,0,0)
\Big].
\end{align}
Here  the l.h.s.  may be interpreted as the one-particle partition functions corresponding to the single sector of the $U(N)$ boundary theory, 
with $\Z_\phi = \Z_{\{0\}}, \ \Z_\psi = \Z_{\{{1\ov 2}\}},\ \Z_T = \Z_{\{1\}}$  given in \rf{26},\rf{a4}.

 The case of the real $O(N)$ invariant theory is found  by an appropriate $\mathbb Z_{2}$ projection. 
The corresponding sums then  represent the 
partition functions of the singlet sector of $O(N)$ invariant free real scalar, Majorana fermion,  and real self-dual tensor theories in 6d:
\begin{align}
\label{C.8}
\hh \big[\Z_{\phi}(q)\big] ^{2}+\hh \Z_{\phi}(q^{2}) &= 
\Z^{+}(4;\,\mathbf{0})+\sum_{s=2,4,\dots}^{\infty}\Z^{+}(4+s;\, s,0,0),\\
\hh \big[\Z_{\psi}(q)\big]^{2}-\hh \Z_{\psi}(q^{2}) &= 
\sum_{s=2,4,\dots}^{\infty}\Z^{+}(4+s;\,s,1,1)
+ \sum_{s=1,3,\dots}^{\infty}\Z^{+}(4+s;\,s,0,0) 
, \label{C.9} \\
\notag
\hh\big[\Z_{T}(q)\big]^{2}+\hh \Z_{T}(q^{2}) &= 
\sum_{s=2,4,\dots}^{\infty} \Big[ \Z^{+}(4+s;\,s,2,2)  +  \Z^{+}(4+s;\,s,0,0)\Big] 
\\
\label{C.10}
&\qquad \qquad  +\sum_{s=3,5,\dots}^{\infty}\Z^{+}(4+s;\,s,1,1)
\ .
\end{align}
These relations \rf{C.5}--\rf{C.7}  can be generalized  by considering a  tensor product of the 
linear  combination of singletons: 
$\big[n_{\phi}\, \{0\}+n_{\psi}\,\{\hh\}+n_{T}\,\{1\}\big]\times \big[n_{\phi}\, \{0\}+n_{\psi}\,\{\hh\}+n_{T}\,\{1\}\big]
$. This gives  for the corresponding characters 
\be
\label{C.12}
\begin{split}
& \Big[ n_\phi\, \Z_\phi (q)  + n_\psi\, \Z_\psi (q) + n_T\, \Z_T(q)  \Big]^2 \\
&\quad  = n_\phi^2\, \sum_{s=0}^{\infty} \Z^+(s+4;s,0,0)
 + n_\psi^2\,  \sum_{s=1}^{\infty} \Big[\Z^+(s+4;s,0,0) + \Z^+(s+4;s,1,1) \Big]\\
&\qquad+n_T^2\, \sum_{s=2}^{\infty} \Big[\Z^+(s+4;s,0,0)+ \Z^+(s+4;s,1,1) + \Z^+(s+4;s,2,2) \Big]\\
&\qquad+2\, n_\phi\, n_\psi\, \sum_{s=0}^{\infty}
\Z^+\left({\textstyle \frac{9}{2}} + s, \hh + s, \hh, \hh\right)
 +2\, n_\phi\, n_T\,\sum_{s=1}^{\infty} \Z^+(s+4;s,1,1) \\
&\qquad+2\, n_\psi\, n_T\,\sum_{s=1}^{\infty} 
\Big[\Z^+\left({\textstyle\frac{9}{2}} + s; \hh + s, \hh, \hh\right) + 
\Z^+\left({\textstyle\frac{9}{2}} + s, \hh + s, {\textstyle \frac{3}{2}, \frac{3}{2}}\right) \Big].
\end{split}
\ee
It is of interest to consider also the  reducible 
case when  the  boundary theory is represented by an  unrestricted  2-tensor, i.e. 
 the parity-invariant  combination  of self-dual and  anti self-dual tensors, i.e. $\{1\}_c= (3;1,1,1) + (3; 1,1,-1)$. 
Then the corresponding  Flato-Fronsdal type relation becomes (cf. \rf{C.4})
\ba
&\{1\}_{c}\times \{1\}_{c} = 2\,\Big[ (6;\,2,2,0)+\,(6;\,1,1,0)+\,(6;\,0,0,0)\Big]  +2\,\bigoplus_{s=3}^{\infty}(4+s;\,s,2,0)\no  \\
&\quad\quad +\bigoplus_{s=2}^{\infty} \Big[
(4+s;\,s,2,2)_c+(4+s;\,s,1,1)_c
+ 2\,(4+s;\,s,0,0)  \Big] \ . \la{c12}
\end{align}
Then  $\Z_{  \{1\}_{c}   }= 2\,\Z_{T}$ and one  finds that \rf{C.7} is replaced by 
\ba
&\big[ 2 \Z_T(q) \big]^2  = 2\,\Big[  \Z^{+}(6;\,2,2,0)+\Z^{+}(6;\,1,1,0)+\Z^{+}(6;\,0,0,0)\Big] + 2\,\bigoplus_{s=3}^{\infty}\Z^+
(4+s;\,s,2,0)\no  \\    
& \qquad   \qquad   +\bigoplus_{s=2}^{\infty} \Big[
\Z^{+}(4+s;\,s,2,2)_c+\Z^{+}(4+s;\,s,1,1)_c
+ 2\Z^{+}(4+s;\,s,0,0)  \Big] \ .   \la{c13}
\end{align}
Also, the analog of \rf{C.10} is 
\begin{align}
\hh \big[2\,\Z_{T}(q)\big]^{2}+\hh \big[2\,\Z_{T}(q^{2})\big]  &= 
\Z^{+}(6;\,2,2,0)+\Z^{+}(6;\,1,1,0)+\Z^{+}(6;\,0,0,0)  \no  \\
&+\sum_{s=2,4,\dots}^{\infty} \Big[
\Z^{+}(4+s;\,s,2,2)_c    + 2\Z^{+}(4+s;\,s,0,0)   \Big]  \no \\  &
+\sum_{s=3,5,\dots} \Z^{+}(4+s;\,s,1,1)_c  +\sum_{s=3}^{\infty}\Z^{+}(4+s;\,s,2,0) \ . \la{c14}
\end{align}

\def \hhh {{1\ov 2}}

\section{Casimir energy for spin $0, \hh, 1$ singletons  in AdS$_{d+1}$}

It is useful to derive the general expressions for the Casimir energy for spin $j=0, \hh,1$  $SO(2,d)$ singletons in the general  case of 
even dimension $d=2,4,6,...$  of the boundary.  

For $j=0,\hh$  the corresponding character or one-particle partition functions are  readily  found, e.g.,  by 
counting states of free scalar or fermion 
in $d$ dimensions \footnote{The singleton with spin $j$ occurs with two possible chiralities.  Here we  consider  one of them.}
\ba\la{f1}
\Z_{0}(q) = \frac{q^{\frac{d-2}{2}}\,(1-q^{2})}{(1-q)^{d}} \ , \ \ \ \ \ \ \ \ \ \ \  
\Z_{\hhh}(q) = 2^{\frac{d}{2}}\, \frac{q^{\frac{d-1}{2}}\,(1-q)}{(1-q)^{d}} 
\end{align}
These satisfy 
\be\la{f2}
\Z_{0}(q) + \Z_{0}(q^{-1})=0\ , \qquad \qquad 
\Z_{1\ov 2 }(q)  +  \Z_{1 \ov 2}(q^{-1})=0 \ .
\ee
As a consequence, in any even $d$  the Casimir energy associated with the $U(N)$ singlet partition functions $[\Z_{0}]^{2}$ and $[\Z_{\hhh}]^{2}$
vanishes because these functions are invariant under $q\to q^{-1}$. 

The character of the $j=1$ singleton representation is \cite{Dolan:2005wy}\footnote{Explicitly,  
we find $\Z_{1}(q) = P_{d}(q)/(1-q)^{d}$ where 
\begin{align}\la{f3}
\notag
P_{4}(q) &= 3 q^2-4 q^3+q^4, 
&
P_{8}(q) &= 35 q^4-56 q^5+28 q^6-8 q^7+q^8,\\
\notag
P_{6}(q) &= 10 q^3-15 q^4+6 q^5-q^6, 
& 
P_{10}(q) &= 126 q^5-210 q^6+120 q^7-45 q^8+10 q^9-q^{10}.
\end{align}
}
\be
\label{6.3}
\Z_{1}(q) = \frac{1}{[(\frac{d}{2}-1)!]^{2}}\,\sum_{n=0}^{\infty}{\te \frac{(n+d-1)!}{n!\,(n+\frac{d}{2})} }\,q^{n+\frac{d}{2}}
 = \frac{d!}{2\,(\frac{d}{2})!^{2}}\,\frac{q^{\frac{d}{2}}}{(1-q)^{d-1}}\,
{}_{2}F_{1}(1, 1-\textstyle\frac{d}{2}; 1+\frac{d}{2}; q) \ . 
\ee
One can check that 
\be\la{f3} 
\Z_{1}(q)+\Z_{1}(q^{-1}) = (-1)^{d/2}\ , 
\ee
so that 
\be\la{f4}
\begin{split}
\big[\Z_{1}(q)\big]^{2} 
=  \hh \Big( \big[ \Z_{1}(q)\big]^{2}+\big[ \Z_{1}(q^{-1})\big] ^{2}- 1 \Big)  +(-1)^{\frac{d}{2}}\,\Z_{1}(q)\ .
\end{split}
\ee
The first  term   in the r.h.s. is  symmetric under $q\to q^{-1}$ and thus  it does not contribute the Casimir energy. 
As a result, we find for the  Casimir energy of the  product of two spin 1 singletons 
(see \rf{2.8},\rf{2.9}) 
\be\la{f6}
E_{c}(\{1\}\times \{1\}) = (-1)^{\frac{d}{2}}\, E_{c}(\{1\})\ .
\ee
For example,  for the  boundary $U(N)$ theory  decsribed by 
the $d=4$ vector  corresponding to  $\{1\}_{\rm c}$   (self-dual and anti self-dual strength) 
we  get  \ci{Beccaria:2014xda}
\be\la{f7}
d=4: \ \ \ \ \   E_{c}\big(\{1\}_{\rm c}\times \{1\}_{\rm c}\big) =4\, E_{c}(\{1\}\times \{1\}) = 4\, E_{c}(\{1\})  = 2\, E_{c}(\{1\}_{\rm c}),
\ee
while for  chiral singleton in 6d, {i.e.} self-dual (or anti self-dual) 6d tensor
\be\la{f8}
d=6: \ \ \ \qquad \  E_{c}(\{1\}\times \{1\}) = -E_{c}(\{1\}),
\ee
in agreement with (\ref{57}),(\ref{59}).

Similar results can be obtained   in the $O(N)$  case   of   real boundary singleton theory, 
  i.e.   for  the  singlet partition functions \ci{Giombi:2014yra,Beccaria:2014zma}
\be\la{f9}
\Z_{j, \text{real}}(q) = \hh \big[\Z_{j}(q)\big]^{2}+\hh (-1)^{2s}\,\Z_{j}(q^{2})\ .
\ee
In this case\foot{Here $n\cdot\{j\}$  denotes $n$ copies of the singleton,  with partition function $n\,\Z_{j}$.
Note also that $E_{c}(\{j\}_c) = 2 E_{c}(\{j\})$.  }
\be\la{f10}
\begin{split}
E_{c}(\{0\}\times\{0\})_{\rm  real } &= E_{c}(\{0\}),\qquad \qquad 
E_{c}(n\cdot\{\hh\}\times n\cdot\{\hh\})_{\rm  real } = n\,E_{c}(\{\hh\}),\\
&E_{c}(n\cdot \{1\}\times n\cdot \{1\})_{\rm  real } = {\te \frac{2+(-1)^{d/2}\,n}{2}}\,E_{c}(\{1\}).
\end{split}
\ee
Then  in 4d   for  the   scalar, Dirac fermion and the vector we recover the results from \ci{Giombi:2014yra,Beccaria:2014zma}
%
\begin{align}
&d=4: \ \ \ \ \   E_{c}(\{0\}\times\{0\})_{\rm  real } = E_{c}(\{0\}),\qquad 
E_{c}(\{\hh\}_{\rm c}\times \{\hh\}_{\rm c})_{\rm  real } 
= 2\,E_{c}(\{\hh\}) = E_{c}(\{\hh\}_{\rm c}),\no  \\
&\qquad \qquad \ \ \ E_{c}( \{1\}_{\rm c}\times  \{1\}_{\rm c})_{\rm  real } 
 = 4\, E_{c}(\{1\}) = 2\, E_{c}(\{1\}_{\rm c}). \la{f11}
\end{align}
In 6d  we get  instead 
\begin{align}
d=6: \ \ \ \  & E_{c}(\{0\}\times\{0\})_{\rm  real } = E_{c}(\{0\}),\qquad
E_{c}(\{\hh\}\times \{\hh\})_{\rm  real } = E_{c}(\{\hh\}),  \no \\
& E_{c}(\{1\}\times  \{1\})_{\rm  real } = \hh \,E_{c}(\{1\}), \la{f12}
\end{align}
in agreement with \rf{5.4},\rf{56},\rf{58}.

\section{Field content of   KK towers in  6d supergravity on $S^3$ } 
\label{app:tower}

Here  we collect the field content of the KK towers discussed in section \ref{sec:ads3}. 
Let us   list the representations of fields  transforming in the $(\Delta; s)\times (j_{1}, j_{2})$
representations of $SO(2,2)\times SO(4)$ as a formal sum of the form 
\be
\label{E.1}
\sum n_{\Delta, s; j_{1}, j_{2}}\,q^{\Delta}\, x^{s}\, R_{j_{1}, j_{2}}\ .
\ee
For states of  the spin-2 tower  in (\ref{5.5})  with   $\ell\ge 0$  we  get  
\begin{align}
&\Phi_{2}(\ell) =\te \big(\frac{\ell+1}{2},\frac{\ell+3}{2}\big)_{\rm s}
+\big(\frac{\ell+3}{2}, \frac{\ell+1}{2}\big)_{\rm s} \no  \\
&= q^{\ell}\,\Big[q^2 \big(x R_{\frac{\ell +1}{2},\frac{\ell
   +3}{2}}+x^{-1}\,R_{\frac{\ell +3}{2},\frac{\ell
   +1}{2}}\big) \no \\
   &\ \ +q^{5/2} \big(2 x^{3/2} R_{\frac{\ell
   +1}{2},\frac{\ell +2}{2}}+2 x^{-3/2}R_{\frac{\ell
   +2}{2},\frac{\ell +1}{2}}+2 x^{1/2}
   R_{\frac{\ell }{2},\frac{\ell +3}{2}}+2x^{-1/2}
   R_{\frac{\ell +3}{2},\frac{\ell
   }{2}}\big) \no \\
   &\ \ +q^3 \big(x^2 R_{\frac{\ell
   +1}{2},\frac{\ell +1}{2}}+x^{-2} R_{\frac{\ell
   +1}{2},\frac{\ell +1}{2}}+4 x R_{\frac{\ell
   }{2},\frac{\ell }{2}+1}+4x^{-1}R_{\frac{\ell
   }{2}+1,\frac{\ell }{2}}+R_{\frac{\ell
   -1}{2},\frac{\ell +3}{2}}+R_{\frac{\ell +3}{2},\frac{\ell
   -1}{2}}\big)\no \\
   &\ \ +q^{7/2} \big(2 x^{3/2} R_{\frac{\ell
   }{2},\frac{\ell +1}{2}}+2x^{-3/2}R_{\frac{\ell
   +1}{2},\frac{\ell }{2}}+2 x^{1/2}
   R_{\frac{\ell -1}{2},\frac{\ell +2}{2}}+2x^{-1/2}
   R_{\frac{\ell +2}{2},\frac{\ell
   -1}{2}}\big)\no \\
   &\ \ +q^4 \big(x R_{\frac{\ell
   -1}{2},\frac{\ell +1}{2}}+x^{-1} R_{\frac{\ell
   +1}{2},\frac{\ell
   -1}{2}}\big)\Big].\label{E.2}
   \end{align}
The massless states at $\ell=-1$  are 
\be
\label{E.3}
\begin{split}
 &(0,1)_{\rm s}  
+(1,0)_{\rm s} \\
&\ \ \ = q \big(x R_{0,1}+ x^{-1} R_{1,0}\big)+q^{3/2} \big(2
   x^{3/2} R_{0,\frac{1}{2}}+2x^{-3/2}
   R_{\frac{1}{2},0}\big)+q^2 \big(x^2
   R_{0,0}+x^{-2} R_{0,0}\big).
      \end{split}
\ee
For the  spin $\hhhh$ tower  (\ref{5.11})  we get 
\ba
& \Phi_{\hhhh}(\ell) =\te  \big(\frac{\ell+1}{2},\frac{\ell+2}{2}\big)_{\rm s}
+\big(\frac{\ell+2}{2}, \frac{\ell+1}{2}\big)_{\rm s} \no \\
&= q^{\ell}\,\Big[q^{3/2} \big(\sqrt{x} R_{\frac{\ell +1}{2},\frac{\ell
   +2}{2}}+x^{-1/2}R_{\frac{\ell +2}{2},\frac{\ell
   +1}{2}}\big)   \label{E.4}\\
   &+q^2 \Big(2 x R_{\frac{\ell
   +1}{2},\frac{\ell +1}{2}}+2x^{-1} R_{\frac{\ell
   +1}{2},\frac{\ell +1}{2}}+2 \big(R_{\frac{\ell
   }{2}+1,\frac{\ell }{2}}+R_{\frac{\ell }{2},\frac{\ell
   }{2}+1}\big)\Big)\no \\
   &+q^{5/2} \Big(x^{3/2} R_{\frac{\ell
   +1}{2},\frac{\ell }{2}}+x^{-3/2} R_{\frac{\ell
   }{2},\frac{\ell +1}{2}}+\sqrt{x} \big(4
   R_{\frac{\ell }{2},\frac{\ell +1}{2}}+R_{\frac{\ell
   +2}{2},\frac{\ell -1}{2}}\big)+x^{-1/2}\big(R_{\frac{\ell
   -1}{2},\frac{\ell +2}{2}}+4 R_{\frac{\ell
   +1}{2},\frac{\ell }{2}}\big)\Big)\no \\
   &+q^3 \big(2 x
   R_{\frac{\ell }{2},\frac{\ell }{2}}+2x^{-1} R_{\frac{\ell
   }{2},\frac{\ell }{2}}+2 \big(R_{\frac{\ell
   -1}{2},\frac{\ell +1}{2}}+R_{\frac{\ell +1}{2},\frac{\ell
   -1}{2}}\big)\big)+q^{7/2} \big(\sqrt{x}
   R_{\frac{\ell -1}{2},\frac{\ell }{2}}+x^{-1/2}R_{\frac{\ell
   }{2},\frac{\ell
   -1}{2}}\big)\Big],\no 
   \end{align}
and its massless part at $\ell=-1$ is 
\be
\label{E.5}
\begin{split}
(0,\hh)_{\rm s} 
+(\hh,0)_{\rm s}=    \sqrt{q} \big(x^{1/2}
   R_{0,\frac{1}{2}}+x^{-1/2}R_{\frac{1}{2},0}\big)+q \big(2 x R_{0,0}+2x^{-1}
   R_{0,0}\big).
\end{split}
\ee
For the spin-1 tower  in (\ref{5.5})  we have for  $\ell\ge 0$
\be
\begin{split}
\Phi_{1}(\ell)&= \te \big(\frac{\ell+2}{2},\frac{\ell+2}{2}\big)_{\rm s}\\
&= q^{\ell}\,\Big[q^2 R_{\frac{\ell +2}{2},\frac{\ell +2}{2}}+q^{5/2}
   \big(2x^{-1/2} R_{\frac{\ell +1}{2},\frac{\ell
   +2}{2}}+2 x^{1/2} R_{\frac{\ell
   +2}{2},\frac{\ell +1}{2}}\big)\\
   &\ \ +q^3 \big(x
   R_{\frac{\ell }{2}+1,\frac{\ell }{2}}+x^{-1}R_{\frac{\ell
   }{2},\frac{\ell }{2}+1}+4 R_{\frac{\ell
   +1}{2},\frac{\ell +1}{2}}\big)\\
   &\ \ +q^{7/2} \big(2x^{-1/2}
   R_{\frac{\ell }{2},\frac{\ell +1}{2}}+2
   x^{1/2} R_{\frac{\ell +1}{2},\frac{\ell }{2}}\big)+q^4
   R_{\frac{\ell }{2},\frac{\ell
   }{2}}\Big].
   \end{split}
\ee
Finally, the extra term in (\ref{5.4}), (\ref{5.7}), and (\ref{5.10})  is 
\be
\begin{split}
 (\hh,\hh)_{\rm s}
= q R_{\frac{1}{2},\frac{1}{2}}+q^{3/2} \big(2x^{-1/2}
   R_{0,\frac{1}{2}}+2 x^{1/2}
   R_{\frac{1}{2},0}\big)+4 q^2
   R_{0,0}\ .
\end{split}
\ee

\def \hah {{1\ov 2}} 

\section{Relation between Casimir energy and 2d central charge computed from $AdS_3$ 
for short $SU(2,2\,|\,1)\times SU(2,2\,|\,1)$ multiplets}
\label{app:cardy}

The short multiplet $(J_{1}, J_{2})_{\rm s}$ of $SU(2,2\,|\,1)\times SU(2,2\,|\,1)$ contains, for generic $j_{1}, j_{2}$,
the following representations $(\Delta; s)_{(j_{1}, j_{2})}$ of $SO(2,2)\times SO(4)$ (see \rf{5.2}):
\begin{align}
\label{F.1}
&(J_{1},J_{2})_{\rm s} = \left(J_1+J_2; J_1-J_2\right)_{\left(J_1,J_2\right)}+2
   \left(J_1+J_2+\hh; J_1-J_2-\hh\right)_{\left(J_1,J_2-\hah\right)}\notag\\
   &+2 \left(J_1+J_2+\hh; J_1-J_2+\hh\right)_{\left(J_1-\hah,J_2\right)}+
   \left(J_1+J_2+1; J_1-J_2-1\right)_{\left(J_1,J_2-1\right)}\notag\\
   &+4   \left(J_1+J_2+1; J_1-J_2\right)_{\left(J_1-\hah,J_2-\hah\right)}+\left(J_1+
   J_2+1; J_1-J_2+1\right)_{\left(J_1-1,J_2\right)}\\
   &+2
   \left(J_1+J_2+\textstyle\frac{3}{2}; J_1-J_2-\hh\right)_{\left(J_1-\hah,J_2-1\right)}
   +2
   \left(J_1+J_2+\textstyle\frac{3}{2}; J_1-J_2+\hh\right)_{\left(J_1-1,J_2-\hah\right)}\notag\\
   &
   +\left(J_1+J_2+2; J_1-J_2\right)_{\left(J_1-1,J_2-1\right)}.\notag
\end{align}
The $S^1$  Casimir energy for a 2d  conformal  field in the $SO(2,2)$ representation $(\Delta; s)$ can be found from 
the partition functions $\Z^{+}$ in (\ref{5.12}) and (\ref{5.13}) and using $E_{c} = -2\,E_{c}^{+}$: 
\be
\label{F.2}
\te E_{c}(\Delta; s) = -\frac{1}{12}(-1)^{2s}\,(\Delta-1)\,\big[2\,(\Delta-1)^{2}-1 \big] \ .
\ee
At the same time, the 2d central charge can be  computed  via  ``dual''  route    as the 
coefficient  of te logarithmic  IR divergence of 1-loop partition function  of the corresponding 
higher spin field  in $AdS_3$  
  \cite{Giombi:2013yva,Giombi:2013fka,Giombi:2014iua};  for a single chiral spin $s$ component it  reads
\be
\label{F.3}
c_{_{\rm AdS_3}} (\Delta; s) = (-1)^{2s}\,(\Delta-1)\,\big[(\Delta-1)^{2}-3s^{2}\big]\  .
\ee
Comparing \rf{F.2} and \rf{F.3}  we  observe  that 
the 2d relation $E_{c} = -\frac{1}{12}\,c$  in \rf{20}
 does not   hold for a single massive field. 
 Nevertheless, this  relation holds 
for a massless field with $\Delta=s$, because 
 \be
\label{F.4}
\begin{split}
E_{c}(s; s)   -E_{c}(s+1; s-1) &=\te  -\frac{1}{12}\,\Big[
c{_{\rm AdS_3}}(s; s)-c{_{\rm AdS_3}}(s+1, s-1)
\Big] \\
&=\te  \frac{1}{12}(-1)^{2s} \,\big[1-6\,s\,(1-s)\big]\ .
\end{split}
\ee
It also holds 
    if we  evaluate the total 
    $E_{c}$ and $c= c_{_{\rm AdS_3}} $ for a short multiplet $(J_{1}, J_{2})_{\rm s}$, i.e. 
\be
\label{F.5}
\te E_{c}[(J_{1}, J_{2})_{\rm s}] = \te -\frac{1}{12}\,c\big[(J_{1}, J_{2})_{\rm s}\big] = -\frac{1}{2}\,(-1)^{2(J_{1}+J_{2})}
\,(J_{1}+J_{2}).
\ee
Different expressions are found when additional shortening occur due to particular 
 low values of $J_{1}$ or $J_{2}$, but we checked that  the
  relation $E_{c} = -\frac{1}{12}c$  always  holds.

\bibliography{6d-Biblio}

\bibliographystyle{JHEP}

\end{document}